\documentclass[twocolumn]{aastex701}

\usepackage{amsmath}

\received{11 May 2026}
\revised{24 June 2026}
\accepted{29 June 2026}
\submitjournal{The Astronomical Journal}

\begin{document}

\title{SAOImageDS9: An Essential Tool for Astronomical Exploration}

\correspondingauthor{Antonella Fruscione}
\author[orcid=0000-0002-6414-3954]{Antonella Fruscione}
\affiliation{Center for Astrophysics $|$ Harvard \& Smithsonian, 
60 Garden St., Cambridge, MA 02138, USA}
\email[show]{afruscione@cfa.harvard.edu}

\author[orcid=0009-0000-5187-673X]{Kenny Glotfelty}
\affiliation{Center for Astrophysics $|$ Harvard \& Smithsonian, 
60 Garden St., Cambridge, MA 02138, USA}
\email{kglotfelty@cfa.harvard.edu} 

\author{William Joye}
\altaffiliation{Retired}
\affiliation{Center for Astrophysics $|$ Harvard \& Smithsonian, 
60 Garden St., Cambridge, MA 02138, USA}
\altaffiliation{Retired}
\email{billjoye@me.com} 

\author[orcid=0000-0002-7093-295X]{Jonathan McDowell}
\altaffiliation{Retired}
\affiliation{Center for Astrophysics $|$ Harvard \& Smithsonian, 
60 Garden St., Cambridge, MA 02138, USA}
\email{planet4589@gmail.com} 
\affiliation{Durham University Space Research Center,  Stockton Rd, Durham DH1 3LE, UK}



\begin{abstract}
SAOImageDS9 (DS9) is an open-source, cross-platform application for the visualization and analysis of astronomical data. Developed at the Smithsonian Astrophysical Observatory, DS9 evolved from an example implementation of reusable imaging components into one of the most widely used astronomical display environments. It has remained useful because it supports many astronomical file formats and coordinate systems, can work directly with event data and image cubes, provides interactive region-based analysis, and can communicate with external tools through command-line and messaging interfaces.
DS9 is used in research, mission operations, and education across a range of wavelengths and at both ground- and space-based observatories. This paper summarizes the historical development of DS9, its principal capabilities, and its impact on the astronomical community, and concludes with an appendix describing the internal architecture that has supported its long-term sustainability.
\end{abstract}

\keywords{\uat{Astronomy software}{1855} ---  \uat{Open source software}{1866} --- \uat{Software documentation}{1869} }

\section{Introduction}
SAOImageDS9\footnote{\url{https://ds9.si.edu}} (hereafter DS9) is a widely used software application for viewing, exploring, and analyzing astronomical data. DS9 is often employed for image display, but it is more accurately described as an interactive analysis environment, combining direct data manipulation, accurate coordinate handling, region-based selection, and interoperability with external analysis tools. Distributed as open-source code and available across Linux, macOS, and Windows, DS9 has become an essential component of global astronomical research, and is used by astronomers, observatory operators, instrument scientists, students, and amateur astronomers.

This versatility comes from combining ease of use with support for detailed analysis. It reads standard astronomical data products such as FITS images and tables\footnote{\url{https://heasarc.gsfc.nasa.gov/docs/heasarc/fits.html}}
(originally \citealt{wells1981}), interprets World Coordinate Systems (WCS; \citealt{greisen2002}), which provide the mathematical mapping between image pixels and celestial coordinates, and supports large datasets efficiently. Beyond its graphical interface, DS9 can be controlled via messaging systems such as the X11 Public Access (XPA; \citealt{joye2000}) protocol and the Simple Application Messaging Protocol (SAMP; \citealt{taylor2012}) a Virtual Observatory (VO)\footnote{\url{https://www.ivoa.net/}}
 standard that enables real-time application-to-application data exchange including control of DS9 from python scripts. Together, these capabilities support both interactive exploration and scripted analysis.

In this paper we review the development of DS9, summarize its principal user-facing capabilities, and discuss its scientific and educational impact. We also highlight the architectural decisions that have allowed DS9 to remain relevant across decades of evolving data standards, astronomy missions, and computing environments. Two appendices provide additional detail on the internal design and code structure.

\section{History of DS9 Development}

DS9 grew out of a lineage of visualization tools developed at the Smithsonian Astrophysical Observatory (SAO) starting in 1990. The original, SAOImage,  was among the first publicly available X11-based astronomical applications and established core scientific display techniques still in use today (\citealt{vanhilst1990, mink1996}). In the mid-1990s, SAOImage: The Next Generation (SAOtng) was developed using IRAF's XIMTOOL graphics libraries \citep{tody1986} and Tool Command Language (Tcl). Tcl provided a high-level scripting environment that allowed for more rapid development and customization than traditional compiled languages. SAOtng also introduced the XPA which allowed external programs to exchange data and control the visualization interface \citep{mandel1994, joye2000}.

Building on this modular philosophy, development of a complete redesign was initiated under a 1997--2000 NASA grant (NAG5-3996) titled \textit{Future Directions for Astronomical Image Display}. This grant supported the SAOTk project, which provided a Tcl/Tk-based framework (utilizing both the Tcl language and its associated graphical user interface (GUI) toolkit, Tk) for reusable components such as display frames, panners, and magnifiers \citep{joye1999, ousterhout2010}. DS9 emerged as the primary implementation of this framework, inheriting its visualization techniques from SAOImage and its modular architecture and messaging-based interoperability from SAOTk and SAOtng. Upon its 1999 release, it was quickly adopted by major high energy astrophysics missions, including NASA’s Chandra X-ray Observatory and ESA’s XMM-Newton, where it replaced older tools and was tailored to specific mission requirements.

The development of DS9 has been defined by its ability to adapt to shifting user requirements and technological standards. As astronomical datasets grew in complexity, DS9 evolved to support multi-extension FITS files, mosaics, and three-dimensional data cubes. These transitions required flexible rendering strategies and advanced memory management to maintain interactivity with large-scale data \citep{joye2004}. Simultaneously, the growth of online archives transformed DS9 from a local tool into one for distributed analysis, allowing users to invoke remote computations and retrieve results for interactive exploration.

Changes in hardware and standards also required substantial updates. Early versions of DS9 were constrained by the 8-bit (256 color) limitations of displays and system-level colormaps. As hardware improved, the application underwent substantial updates to support 24-bit ``True color" rendering, with 8 bits for each Red-Green-Blue (RGB) color channel, and sophisticated visualization modes. Coordinate handling similarly expanded; while early versions focused on simple projections, DS9 was updated to interpret complex WCS standards, including all-sky representations and data based on Hierarchical Equal Area isoLatitude Pixelization (HEALPix; \citealt{gorski2005}).

In parallel, DS9 transitioned across computing environments, moving from support for specialized Silicon Graphics workstations from the '90s to widely used operating systems such as Linux, macOS, and Windows. Staying adaptable has allowed DS9 to remain compatible with evolving FITS standards and modern operating systems and has been critical to its role as a general-purpose visualization environment. Because DS9 has continued to support accurate astronomical data analysis through many changes in computing, it remains useful for both mission pipelines and multiwavelength research.

\begin{figure}[ht!]
\includegraphics[width=0.47\textwidth]{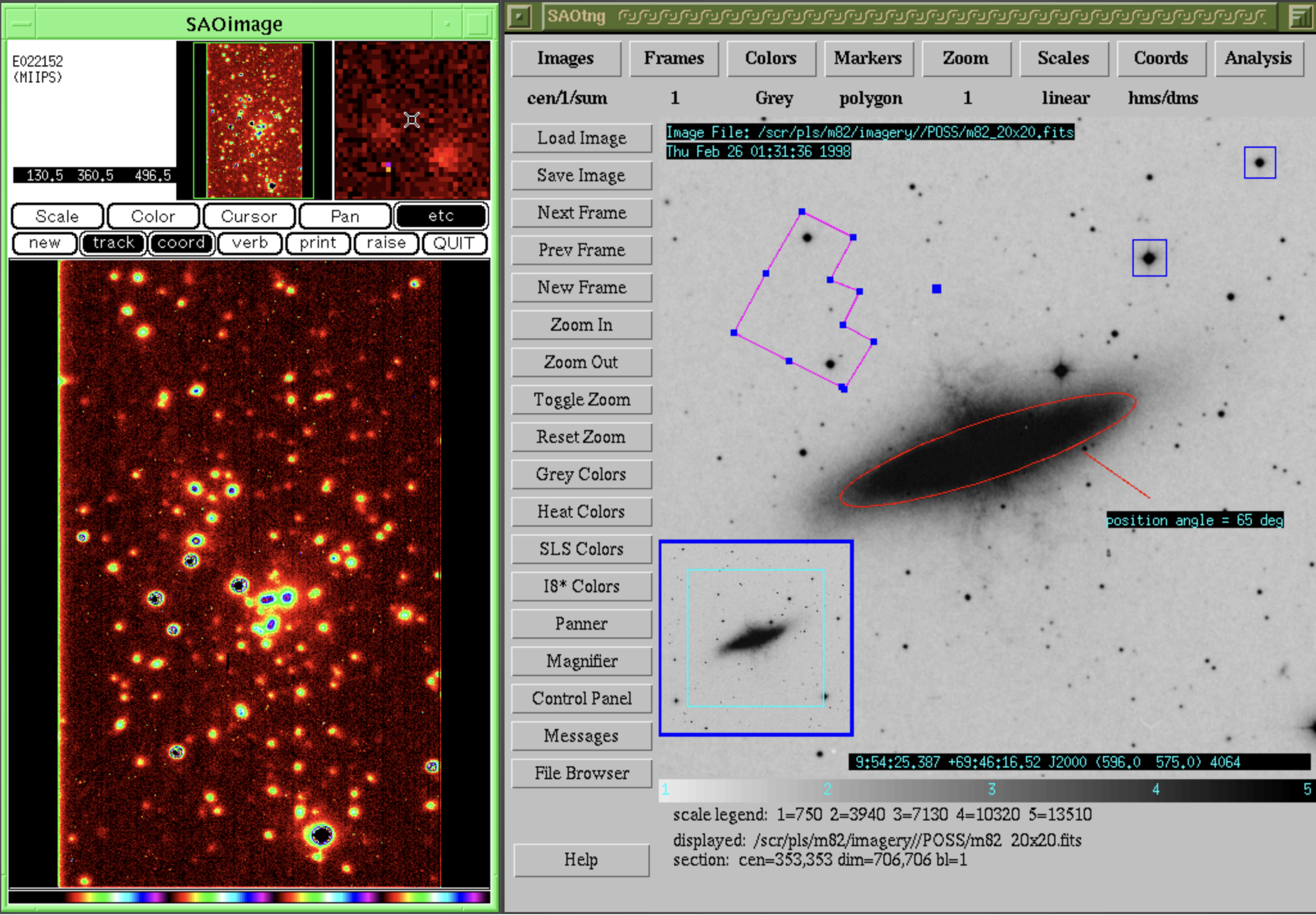}
\caption{Evolution of SAO visualization tools. Left: SAOImage (circa 1990), one of the first X11-based astronomical image display applications, demonstrating early interactive visualization capabilities.  Credit: D. Gudehus, Georgia State University \footnote{\url{https://www.astro.gsu.edu/~gudehus/saoimage_pseudocolor.html}} . Right: SAOtng (mid-1990s), which introduced enhanced graphical interfaces, region handling, and external analysis control via XPA. Credit: P. Shopbell, Caltech\footnote{\url{https://sites.astro.caltech.edu/~shopbell/papers/saotng/}}.}
\label{fig:saoimage}
\end{figure}

\section{Design Principles}

DS9 has been developed with an emphasis on maintaining a consistent structure, ease of use, and flexibility in its implementation \citep{joye2005}. A key feature of DS9 is that graphical interaction and programmatic control are treated as complementary modes. By ensuring that visualization components can be driven both by direct user input and by external messaging interfaces, DS9 provides a unified environment that supports both exploratory "quick-look" analysis and complex automated scripting \citep{joye1999, joye2000}.

DS9 uses a hybrid implementation: a flexible Tcl/Tk graphical interface coupled with compiled C/C++ components for high-performance computational tasks \citep{joye2011}. This structure supports incremental development driven by user requirements rather than rigid, long-term specifications. This approach has allowed DS9 to evolve with changing scientific needs while keeping the user experience relatively stable.

A primary design goal is to minimize the configuration required by the user. DS9 automatically handles technical tasks such as FITS header interpretation, coordinate transformations, and dynamic image scaling. This approach makes DS9 accessible to students and occasional users, while still supporting the detailed analysis needed for research \citep{mandel2001}. Furthermore, DS9 treats regions not just as graphical overlays, but as portable analysis objects. This allows interactive selections to be  integrated into downstream analysis, connecting visual exploration and quantitative results.

\section{Features and Capabilities}
\subsection{Data Visualization}

DS9 provides a visualization environment optimized for astronomical data products, especially FITS images and tables, while also supporting other common bitmap image formats (e.g., PNG, JPEG and TIFF). It can open multi-extension FITS files, display images derived from event files (a tabular format), and work with three-dimensional data cubes. It supports interactive exploration of slices and projections, and interpret both spatial and non-spatial WCS. 
Its display model distinguishes among the full dataset, the portion resident in memory, and the region currently rendered, helping maintain interactivity for large images. Users can zoom, pan, rotate, crop, block, and bin data, and overlay coordinate grids and alternate WCS descriptions.

\begin{figure}[ht!]
\includegraphics[width=0.47\textwidth]{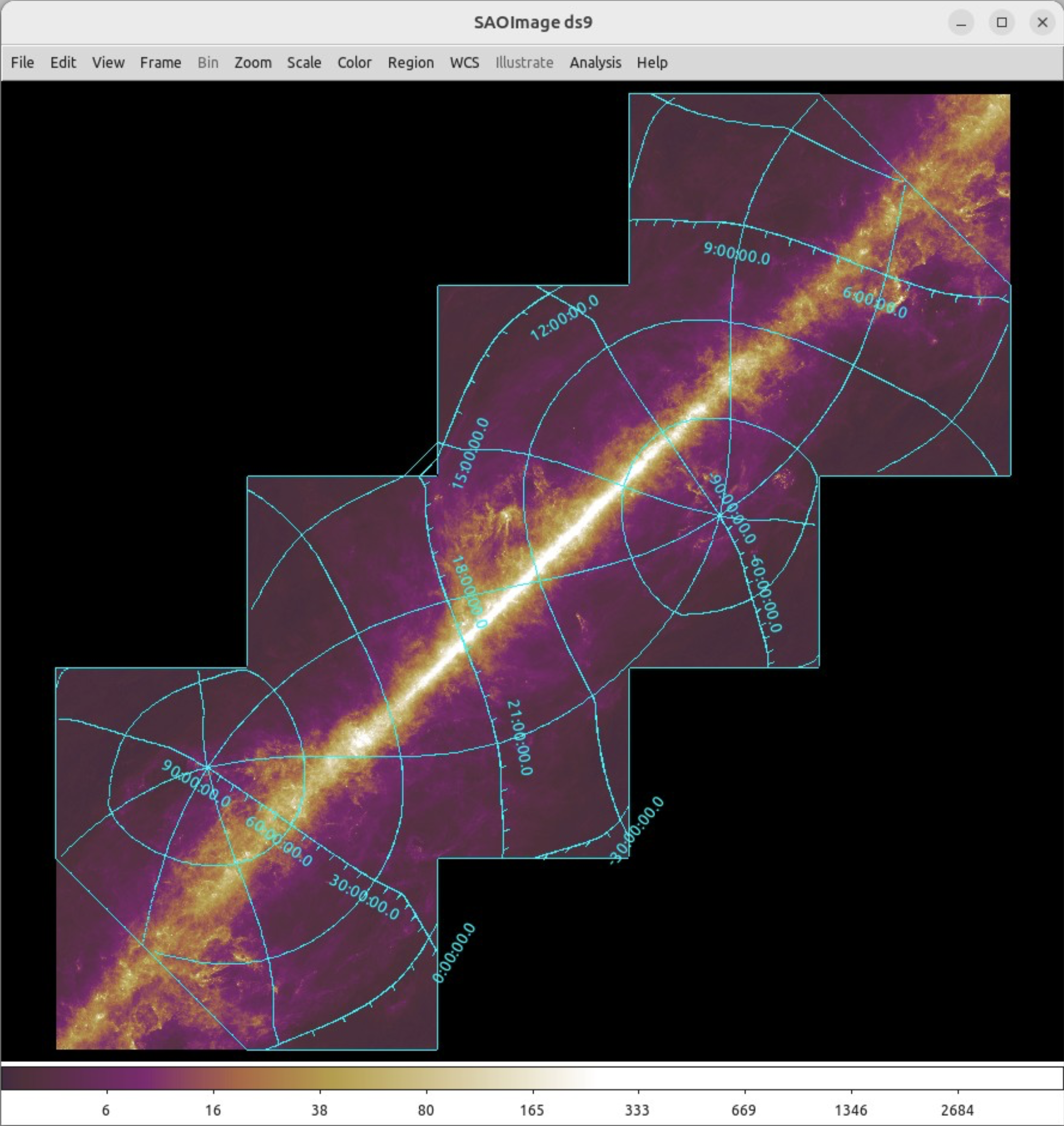}
\caption{All-sky cosmic microwave background from the Planck I-Stokes HEALPix FITS map\footnote{\url{https://irsa.ipac.caltech.edu/data/Planck/release_2/all-sky-maps/index.html}}. The display utilizes a mosaic of HEALPix tiles with overlaid coordinate grids in multiple projections, demonstrating DS9’s ability to handle hierarchical pixelizations and non-standard WCS representations.}
\label{fig:healpix}
\end{figure}

The graphical interface is highly configurable, allowing users to tailor display elements such as coordinate readouts, panners, and magnifiers. The ability to maximize the image display area, combined with native support for specialized formats such as HEALPix (Fig.\ref{fig:healpix}), makes DS9 an effective tool for exploring both high-resolution localized imaging and all-sky survey data \citep{joye2011}.

\subsection{Interactive Analysis}

\begin{figure}[ht!]
\includegraphics[width=0.47\textwidth]{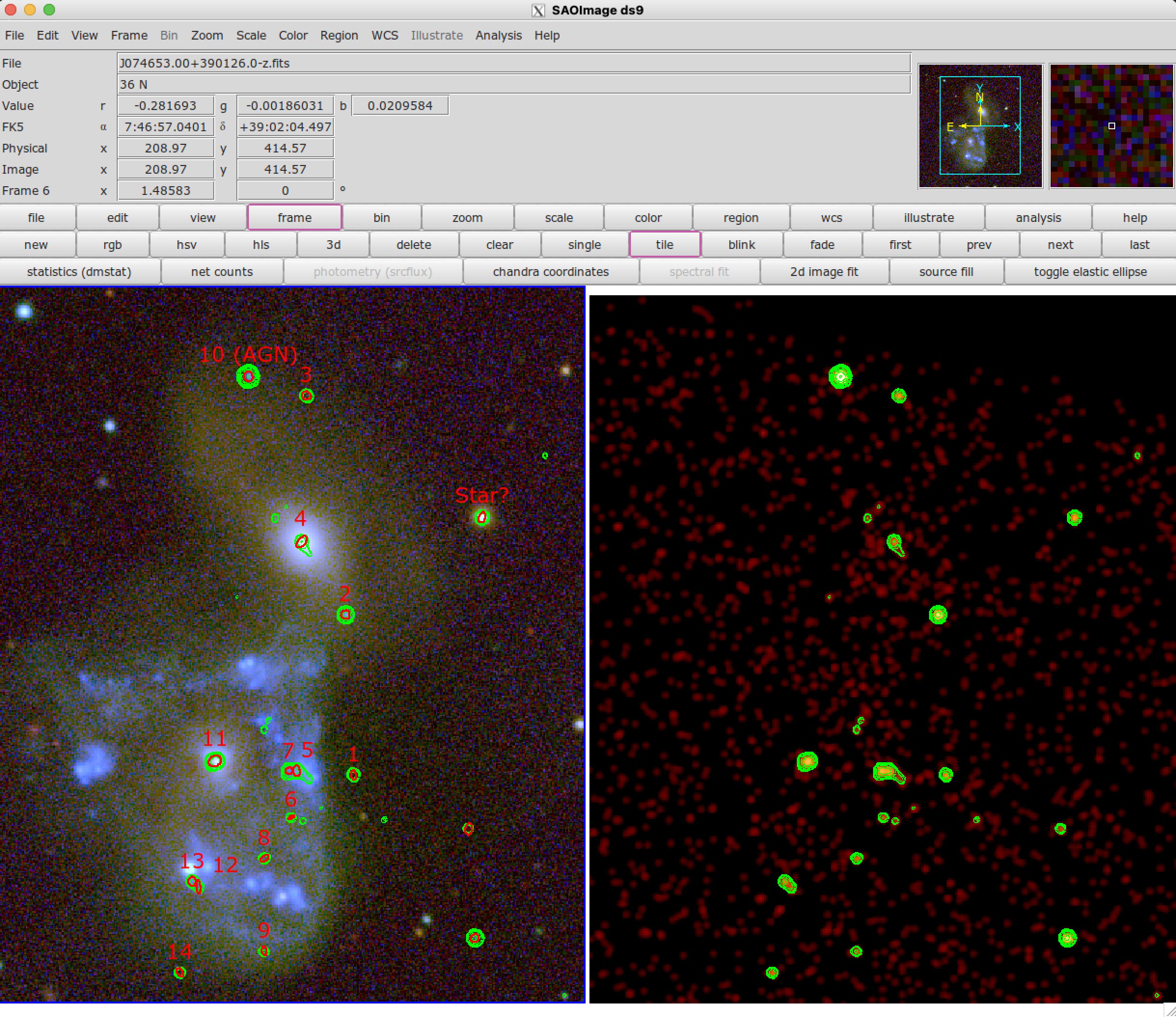}
\caption{Multi-wavelength visualization of the ring galaxy Arp 143 (adapted from \citealt{wolter2018}). (Left) SDSS-III RGB composite (z, g, u bands) with Chandra X-ray sources (red) and contours (green) overlaid. (Right) Smoothed Chandra X-ray event file with intensity contours. The matching frames illustrate coordinate-based synchronization of heterogeneous datasets.}
\label{fig:multiwave}
\end{figure}

A primary use of DS9 is the comparison of heterogeneous datasets through multi-frame display modes (see Figure ~\ref{fig:multiwave}). Images can be tiled, blinked, matched by scale, or aligned in coordinate system. These capabilities support comparison across wavelengths or processing levels and the construction of RGB composites and mosaics. Interactive analysis tools include region creation and editing, centroiding, summation, axis projections, contours, smoothing, and plotting. Regions can be saved, exchanged in different coordinate systems, and used in downstream analysis.

\begin{figure}[ht!]
\includegraphics[width=0.47\textwidth]{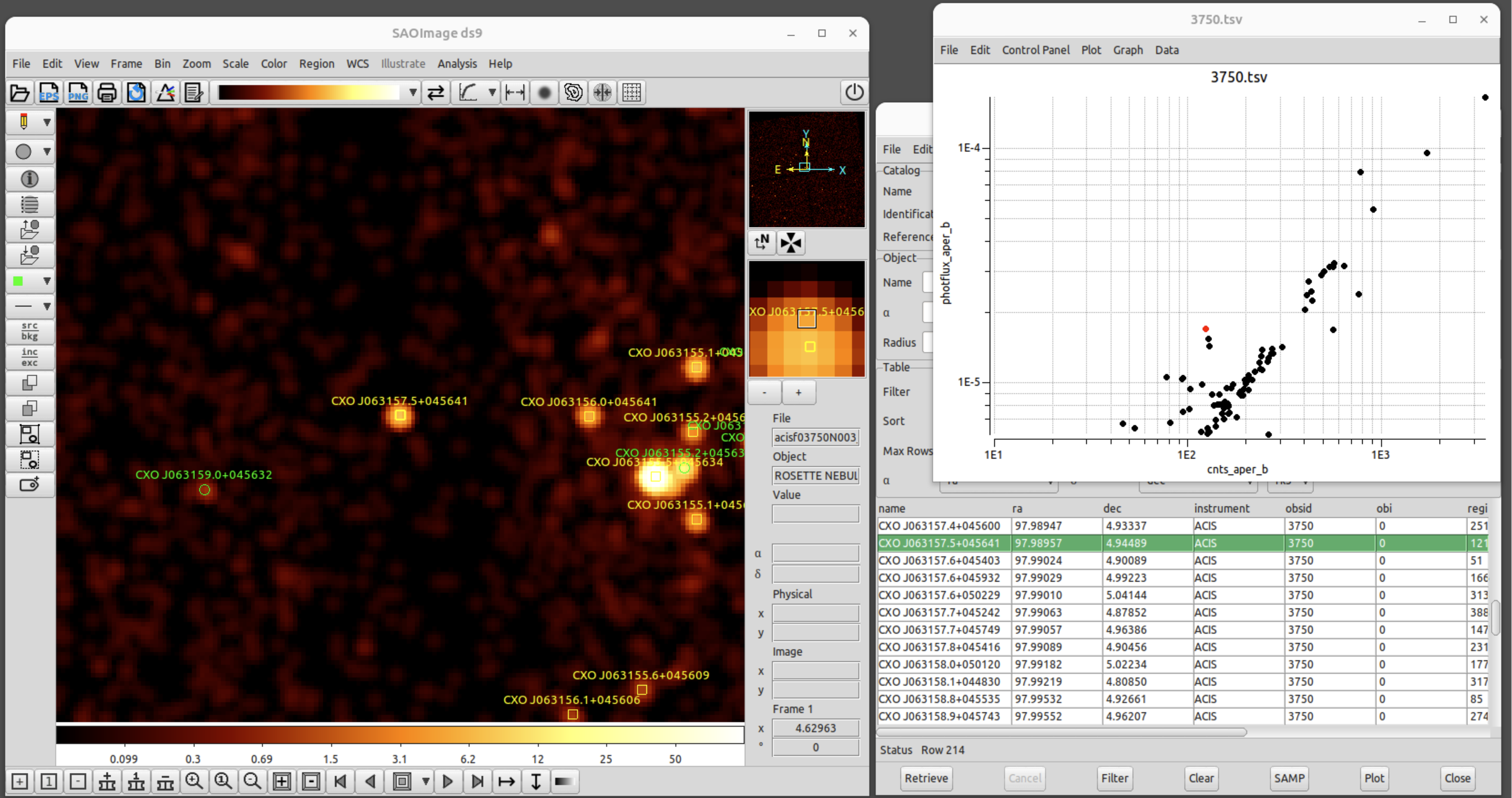}
\caption{Linked-view analysis of the Chandra Source Catalog, shown here utilizing the ``advanced view" layout. Selections made within the X-ray image, scatter plot, or data table are automatically synchronized and highlighted across all views. Variable sources are marked in green and non-variable sources in yellow.}
\label{fig:catalog}
\end{figure}

DS9 also supports tabular data exploration through its catalog tool. VOTables\citep{ochsenbein2025} and FITS tables, including outputs from source detection algorithms and external tools, can be loaded and displayed alongside image data, providing a linked-view approach where selections in a scatter plot or a table automatically highlight corresponding sources in the image (Fig.~\ref{fig:catalog}). This feature enables coordinated exploration of images, catalogs, and derived quantities. \citep{glotfelty2024}.

\subsection{Interoperability}\label{sec:interoperability}

A distinguishing feature of DS9 is that its functionality is accessible through multiple interfaces, including the graphical interface, command-line input, and messaging protocols such as XPA and SAMP. XPA enables external programs to control DS9 and exchange data in real time \citep{joye2000}, while SAMP provides interoperability with other astronomy applications within the VO. This interoperability operates at three distinct levels:

\begin{figure}[ht!]
\includegraphics[width=0.47\textwidth]{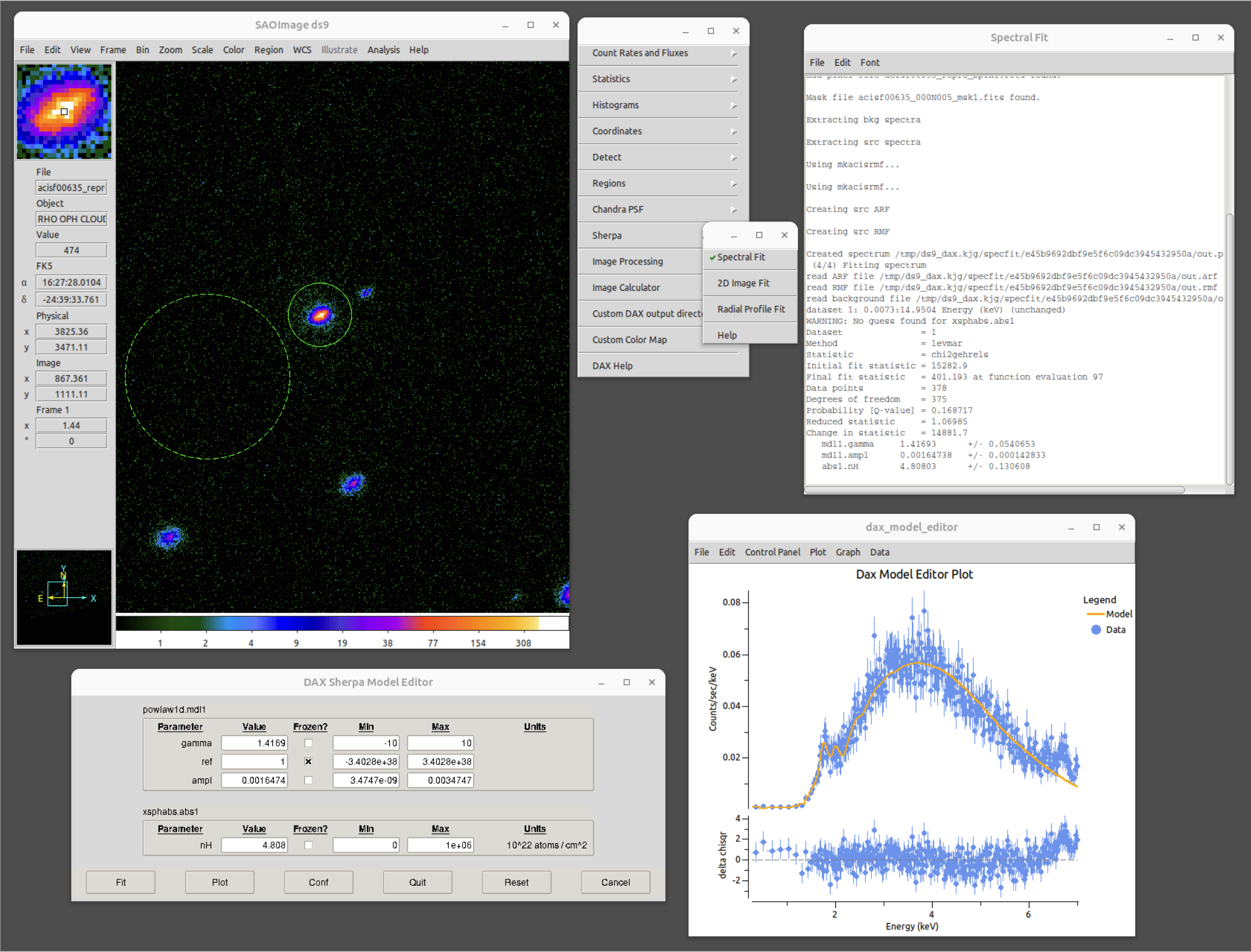}
\caption{Integration of DS9 with Chandra’s CIAO software and the DAX extension. The interface enables seamless spectral extraction and fitting with Sherpa \citep{siemiginowska2024} directly from selected regions (source and background). The Model Editor (lower-left) allows for interactive parameter adjustment, with residuals and best-fit models displayed in real time (lower-right).
}
\label{fig:dax}
\end{figure}

\begin{figure*}[ht!]
\includegraphics[width=1\textwidth]{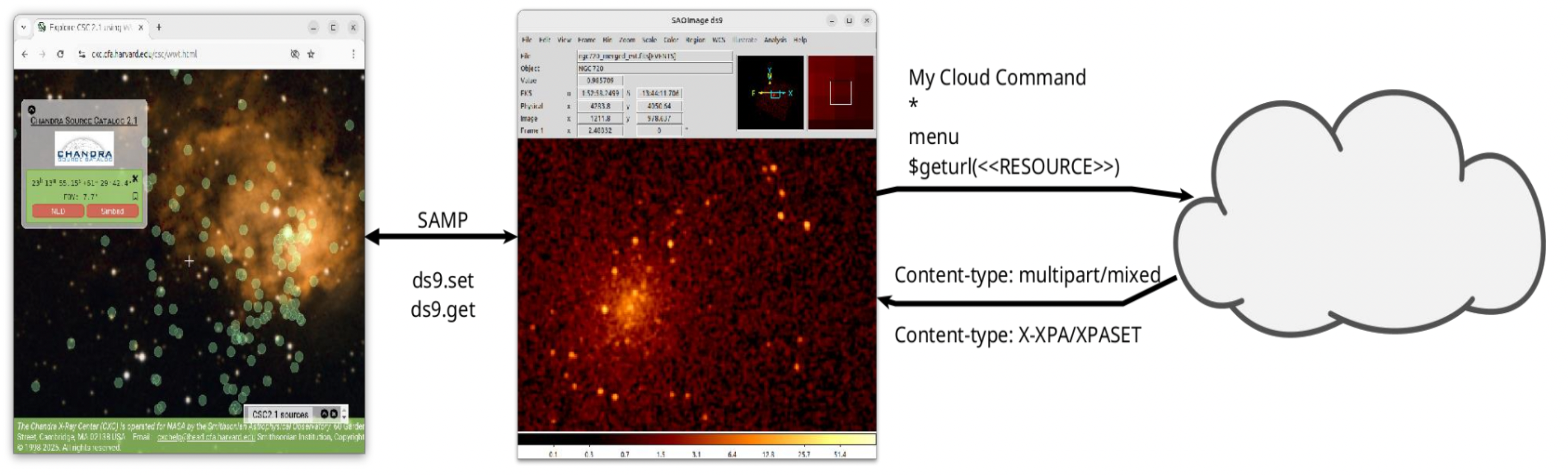}
\caption{Representation of DS9 as a bridge between local, web, and cloud-based environments. (Left) Through the SAMP protocol, the browser based Chandra Source Catalog World Wide Web (WWT) interface communicates with the local DS9 application, allowing the user to select sources in a browser and see them instantly highlighted in the image display. (Center) DS9 provides a SAMP hub through with other applications and browsers can communicate. (Right) DS9 uses HTTP and XPA requests to use remote cloud services for computational tasks. The remote server performs the analysis and sends the results back—either as complex data packages or direct display commands—updating the user's view in real time without the need for local processing power.}
\label{fig:interoperability}
\end{figure*}

\textit{Analysis Level Integration:} DS9 provides a framework for executing external tasks directly within its interface, utilizing user-defined parameters such as regions, coordinates, and data subsets. This is best exemplified by the DS9 Analysis Extension (DAX), developed for the Chandra Interactive Analysis of Observations (CIAO) software package \citep{glotfelty2011, fruscione2026}. DAX allows users to perform spectral extraction, image processing, and model fitting (see Figure~\ref{fig:dax}) using CIAO tools without leaving the visualization environment \citep{glotfelty2020}. Similar integration is available through the \texttt{pyds9plugin}, which enables Python-based quick-look processing \citep{picouet2024}.

\textit{Remote Data and Services:} A second level of interoperability is achieved through support for remote data access. Using built-in HTTP-based mechanisms, DS9 can retrieve data products from remote archives and services, incorporating them directly into the local visualization environment. This capability has proven particularly effective in educational settings. Where relevant, DS9 makes use of VO standards to discover and access remote resources. This includes accessing a curated list of Simple Image Access (SIA)\footnote{\url{https://www.ivoa.net/documents/SIA/}} protocol servers for image retrieval. Currently in development, DS9 will also support VO standard Multi-Order Coverage (MOC)\footnote{\url{https://www.ivoa.net/documents/MOC/}} maps.

\textit{Virtual Observatory Integration:} The third level involves integration with the VO through protocols such as SAMP \citep{taylor2012}. DS9 supports various VO standards, including VOTable catalogs and footprint services, allowing it to interoperate with external applications such as Aladin and TOPCAT \citep{bonnarel2000, baumann2022, taylor2005}. By acting as a SAMP hub or client, DS9 synchronizes its display with external web-based archives and visualization tools, as illustrated in the schematic overview in Figure~\ref{fig:interoperability}. 
Integration of other applicable VO standards will be considered in future updates. Current standards such as the Table Access Protocol (TAP)\footnote{\url{https://www.ivoa.net/documents/TAP/}} are much more suited to applications such as Topcat, and formats such as Hierarchical Progressive Survey (HiPS)\footnote{\url{https://www.ivoa.net/documents/HiPS/}} are better handled by dedicated survey tools such as Aladin.

\subsection{From analysis to publication}
\begin{figure}[ht!]
\includegraphics[width=0.47\textwidth]{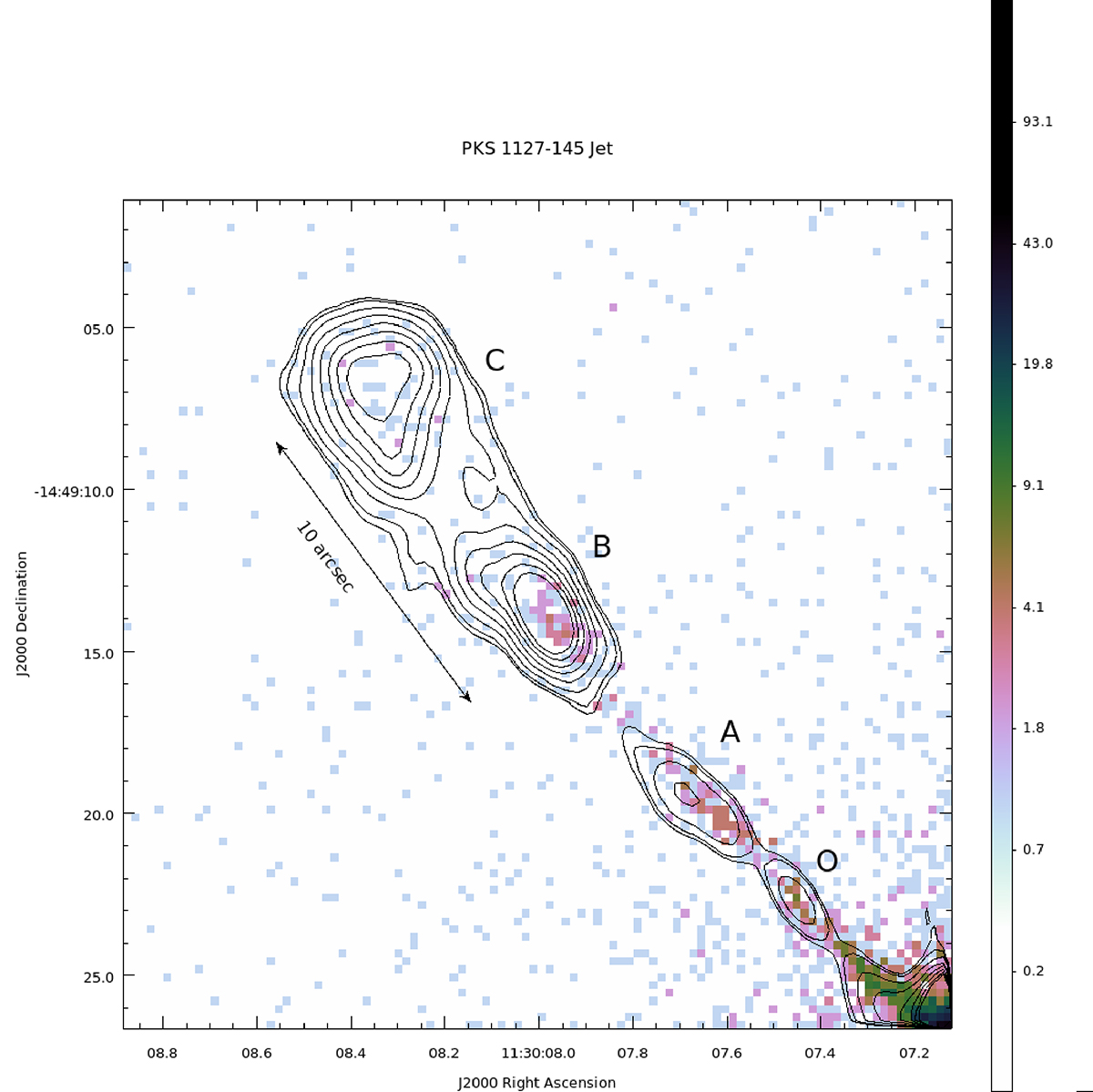}
\caption{Publication-quality multi-wavelength overlay of the PKS 1127-145 jet (adapted from \citealt{orienti2024}). Chandra X-ray data (colored pixels) are aligned with 1.6 GHz VLA radio contours (black lines). DS9’s annotation tools enable precise placement of coordinates, scale bars, and labels directly on the scientific data.}
\label{fig:paper}
\end{figure}

Beyond its use in analysis, DS9 also provides rendering, layering, and annotation tools that support the creation of publication-quality figures. DS9's ability to handle complex overlays, multi-scale contours, and coordinate grids ensures that final images remain scientifically rigorous while meeting the standards of peer-reviewed literature (see Figure~\ref{fig:paper}).

\section{Selected Advanced Features}

Describing all the features and capabilities of DS9 is beyond the scope of this paper, but we are highlighting here a few of the more advanced features useful for scientific analysis.

\begin{figure}[h]
\includegraphics[width=0.47\textwidth]{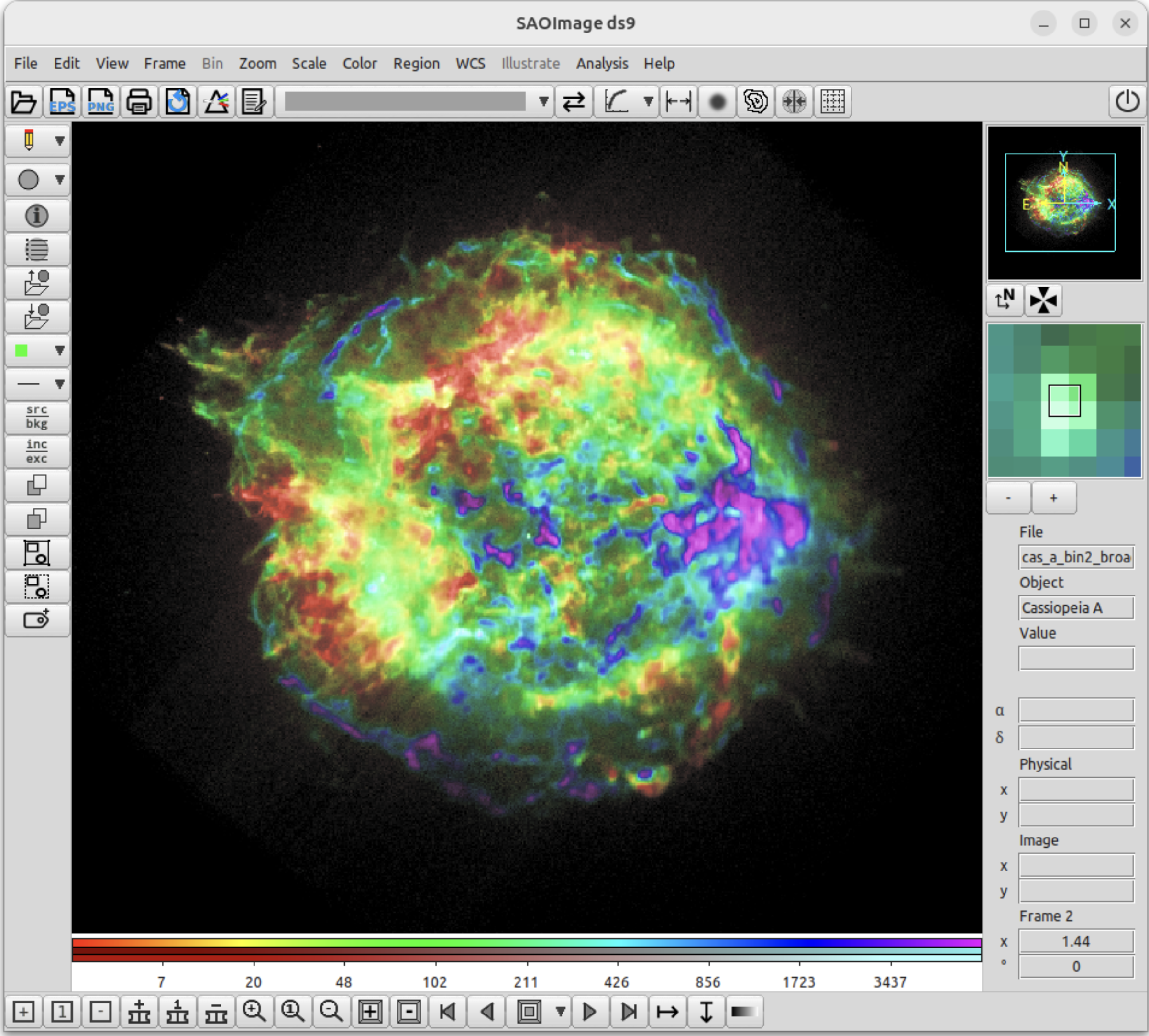}
\caption{HLS representation of Cassiopeia A from Chandra ACIS data. The color hue (red through violet) encodes the mean energy of the events in each pixel, while the exposure-corrected counts are mapped to the lightness and saturation channels. Brighter regions correspond to higher count rates and appear lighter and less saturated, whereas fainter emission appears darker, allowing both the spectral variations and the spatial structure of the remnant to be visualized simultaneously.}
\label{fig:HLS}
\end{figure}

\textit{Tri-Color and True-Color Images} Creating tri-color RGB images in DS9 has been possible for a long time and remains a standard technique, where three different datasets are mapped to the red, green, and blue channels to highlight their relative contributions. DS9 automatically handles the alignment based on the images WCS. In contrast, the use of the Hue-Saturation-Value (HSV) and Hue-Lightness-Saturation (HLS) color models is a more recent addition, offering a different approach in which a single dataset is mapped into a continuous color space.  While RGB emphasizes comparisons between multiple datasets through color mixing, HSV and HLS focus on representing the structure within a single dataset, with hue encoding intensity and brightness controlled through value or lightness (see Figure~\ref{fig:HLS}). This newer capability provides more flexibility in visualizing subtle features, complementing the traditional RGB approach rather than replacing it.

\begin{figure}[h]
\includegraphics[width=0.47\textwidth]{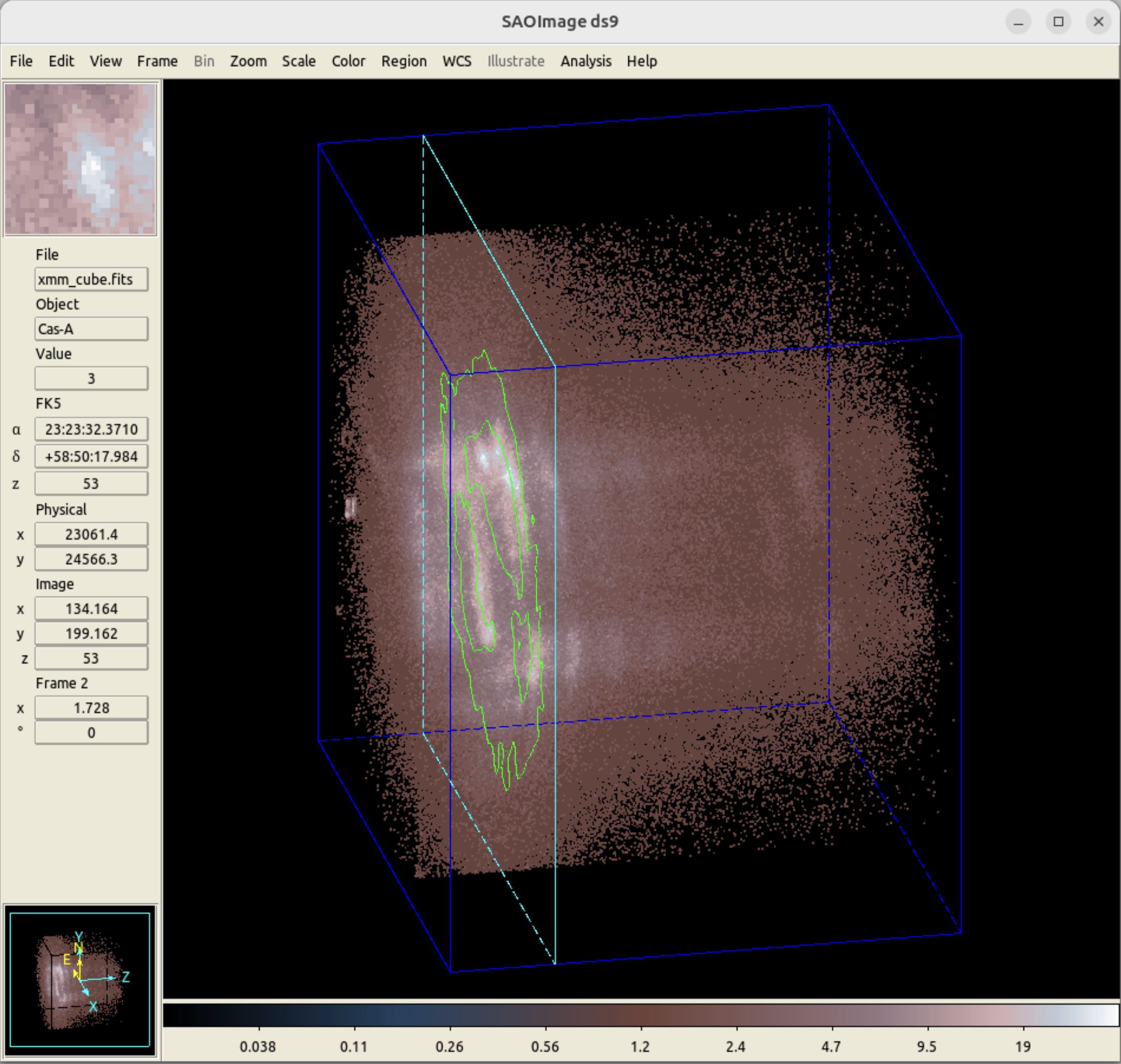}
\caption{Visualization of a 3D XMM-Newton X-ray data cube. The volume rendering shows the distribution of emission, with a selected slice through the cube displayed as a plane. Contours derived from the data are overlaid on the slice. This functionality enables interactive exploration of spatial and spectral structures in multi-dimensional datasets.}
\label{fig:cube}
\end{figure}

\textit{Data Cubes and Slices} In DS9, a data cube is a multi-dimensional dataset that may include spatial, spectral, temporal, polarization, or other axes. DS9 can handle datasets with up to nine dimensions, a capability particularly important in radio astronomy. In many applications, however, the most common case is the three-dimensional cube, with two spatial axes and a third axis representing quantities such as energy, wavelength, or time. These 3-D cubes can be explored by stepping through individual slices, animating them, or collapsing the cube along the third dimension to create integrated images. This allows users to connect spatial structures with their spectral or temporal behavior and to extract quantities like spectra or light curves from selected regions, providing a more complete view of the data than a single static image (see Figure~\ref{fig:cube}).

\begin{figure}[h!]
\includegraphics[width=0.47\textwidth]{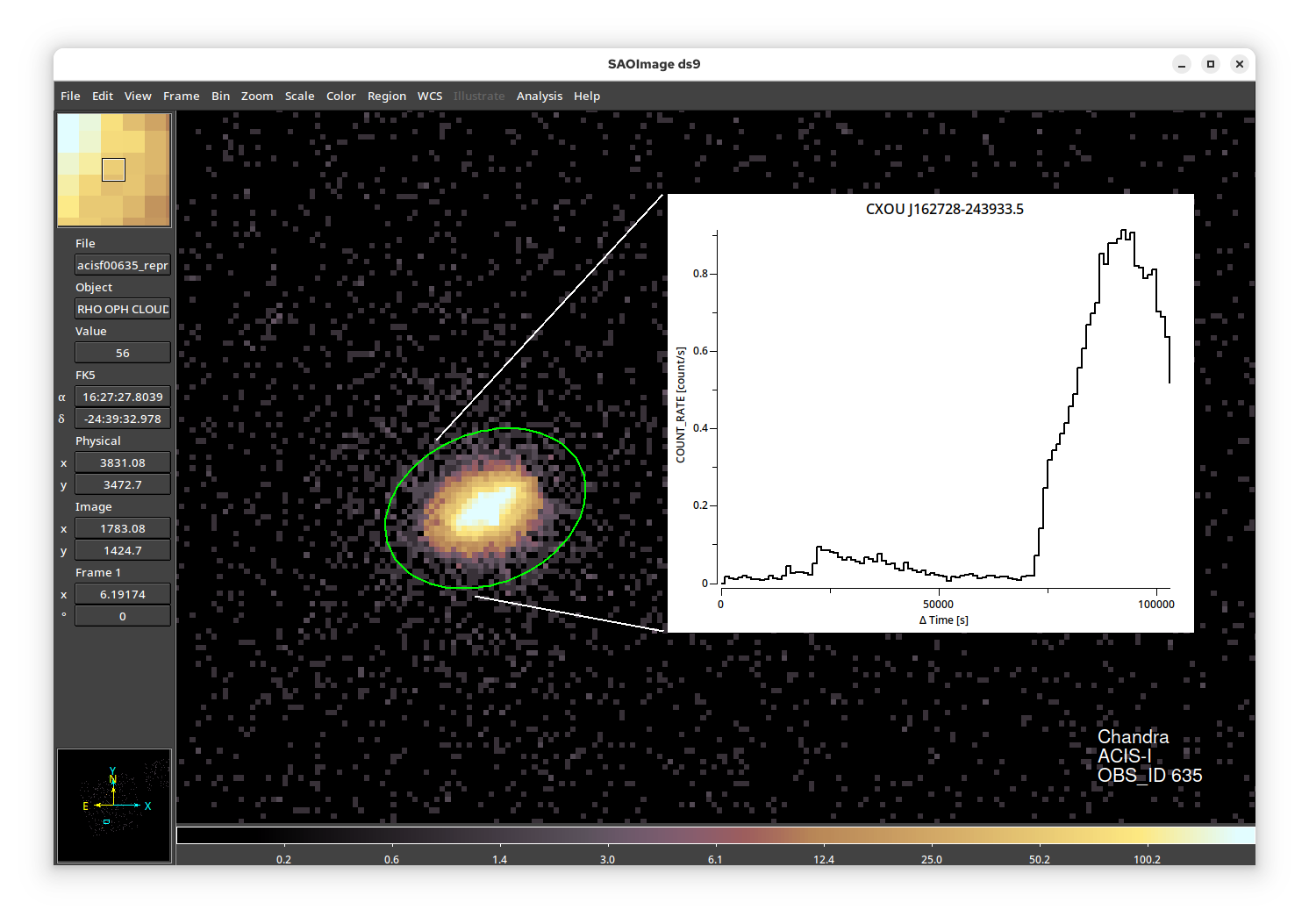}
\caption{Illustrate mode allows users to place annotations in the image display area that remain fixed when panning or zooming. This example shows a Chandra image of a variable source. The source aperture is drawn as a region: green ellipse. The inset light curve plot and the call-out lines are drawn using Illustrate mode.
In this example, the plot itself was also created with DS9. The plot parameters were adjusted using the \textit{Plot Control Center} and then exported to PNG format. The PNG file was then used as an annotation in Illustrate mode. The multi-line text in the bottom right corner created in Illustrate mode.
}
\label{fig:illustrate}
\end{figure}

\textit{Illustrate Mode} allows users to add annotations (text, images, lines, circles, boxes, ellipses, etc.) to images. Unlike regions which are located within the coordinate system of the image, these annotations remain fixed within the display window so as the image is panned or zoom they do not move. This can be used to embed logos in displays, or as is shown in Figure~\ref{fig:illustrate} allows users to add call-out boxes with plots (or any other static bitmap image).

\begin{figure}
\begin{interactive}{animation}{ds9_dart.mp4}
\includegraphics[width=0.47\textwidth]{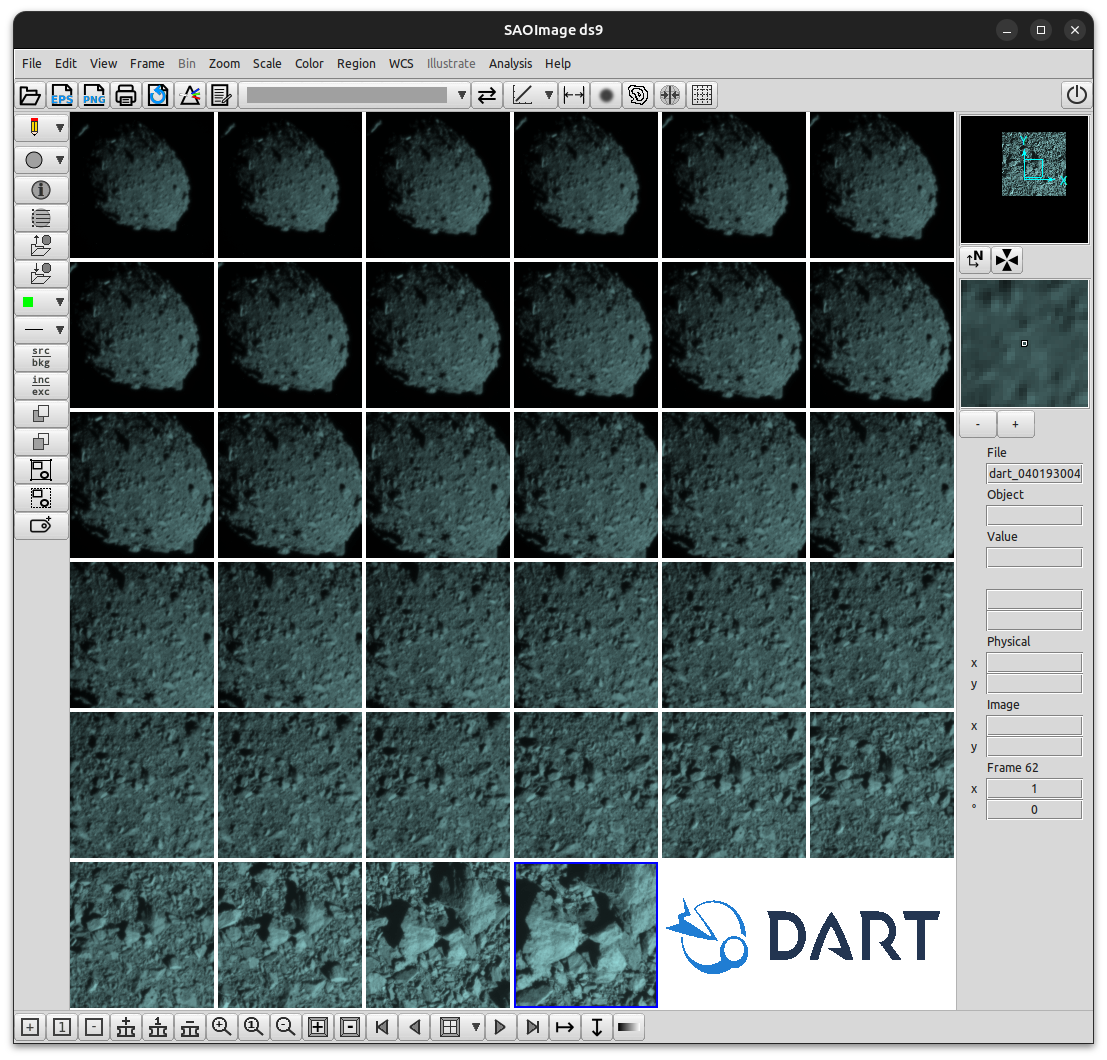}    
\end{interactive}
\caption{This image shows 34 individual FITS images obtained from the DART mission archive that have been manually registered to have the same location on the asteroid located in the center of the image. \textbf{This image links to a three (3) second video} created with DS9 stepping through the individual frames. The video shows the DART spacecraft approaching Dimorphos with the last frame just before impact. The DART mission logo was added in Illustrate mode.}
\label{fig:animation}
\end{figure}

\textit{Animations}. DS9 allows users to directly create animations. These can be generated from multiple individual frames as shown in the video linked to Figure~\ref{fig:animation} which displays the FITS images from the Double Asteroid Redirection Test (DART) mission's impact. Other options include generating animations from individual slices through a multi-dimension data set or rotating the rendering of  3-dimensional about azimuth, elevation, and zoom. Animations can be saved in MPEG format or as animated GIFs.

\begin{figure}
\begin{interactive}{animation}{ellipses.mp4}
\includegraphics[width=0.47\textwidth]{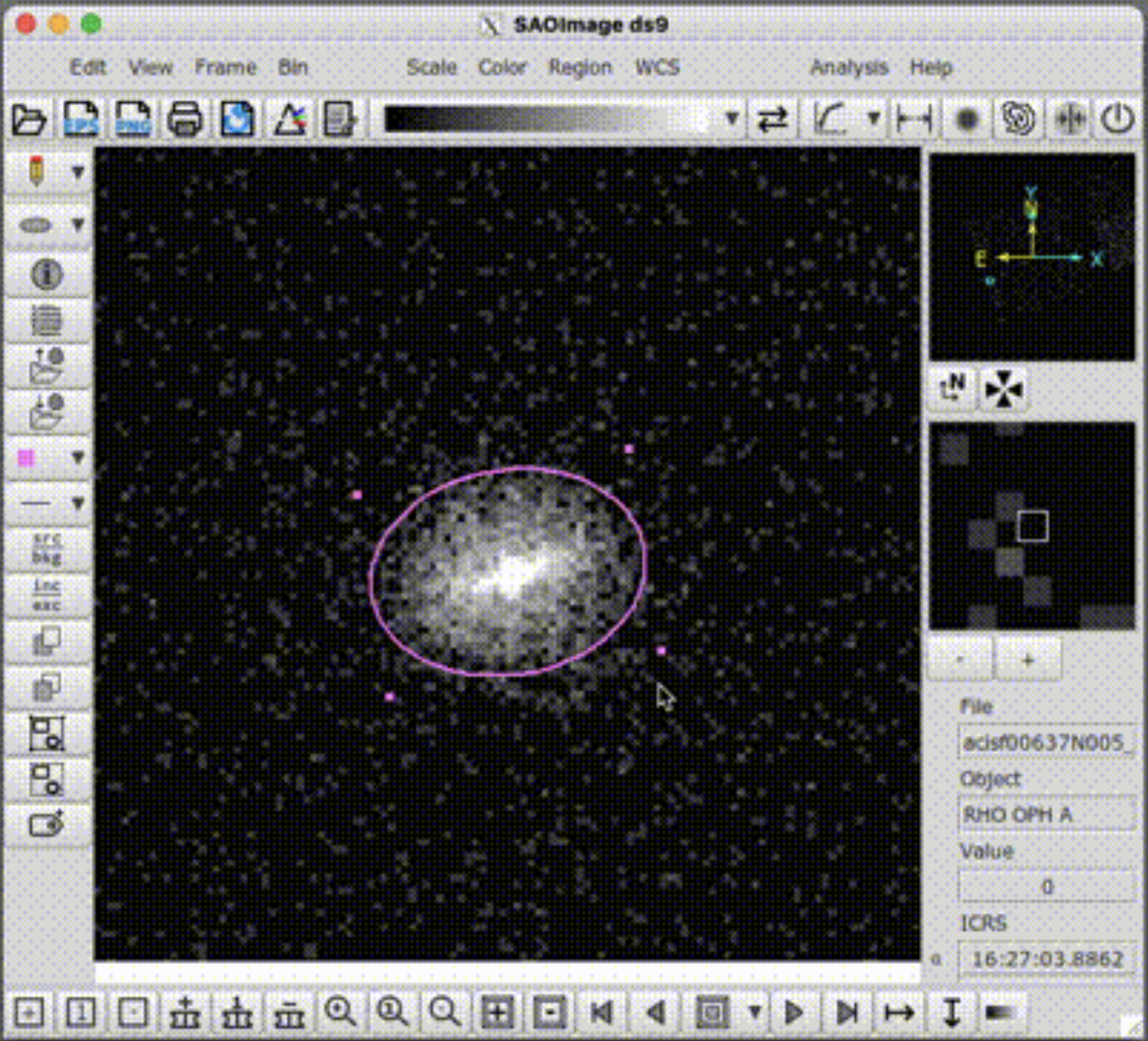}    
\end{interactive}
\caption{Example of the DAX elastic ellipse task applied to an off-axis Chandra source. When activated, the region dynamically resizes, repositions, and reorients to match the source morphology based on the second-order image moments. \textbf{This image links to a twenty-three (23) second video} showing the steps to activate the task: select an elliptical region $\rightarrow$ position it loosely near the source of interest $\rightarrow$ from the Analysis Menu choose CIAO Tools (DAX) $\rightarrow$ Interactive Regions $\rightarrow$ Enable Elastic Ellipses. The ellipse then automatically resizes to the match the size, location, and angle of the data contained in it.}
\label{fig:ellipses}
\end{figure}

\textit{Interactive Regions (DAX)}. DS9 includes advanced region-based tools that support interactive, data-driven analysis. Through the DAX extension, which is implemented within the core DS9 architecture, certain region types can adapt dynamically to the underlying data. One example is the ``elastic ellipse'' task, in which an elliptical region automatically resizes, repositions, and reorients to match the local source morphology based on second-order image moments. As shown in the video linked to Figure~\ref{fig:ellipses}, this provides an intuitive and quantitative way to track extended or asymmetric sources.

\section{DS9 in the community}

\subsection{DS9 and Chandra}

For over two decades, DS9 has been closely linked to Chandra data analysis, mission operations, and observation planning. Currently funded and maintained by the Chandra X-ray Center, DS9 is distributed as a core element of the CIAO software suite. It is utilized throughout the mission lifecycle, from the automated verification and validation pipelines used in quality assurance to the interactive analysis performed by individual researchers.

DAX (see Section~\ref{sec:interoperability}) provides the most direct link between the two, acting as the main bridge between DS9 and the CIAO analysis tools. Within the Chandra workflow, DAX is routinely used to connect interactive source selection with downstream analysis, enabling a seamless transition from visualization to quantitative modeling (see Figure~\ref{fig:dax}). Its tight integration with DS9 allows users to incorporate region definitions, coordinate information, and data filtering directly into CIAO tasks, streamlining the analysis process.

\begin{figure}[ht!]
\includegraphics[width=0.47\textwidth]{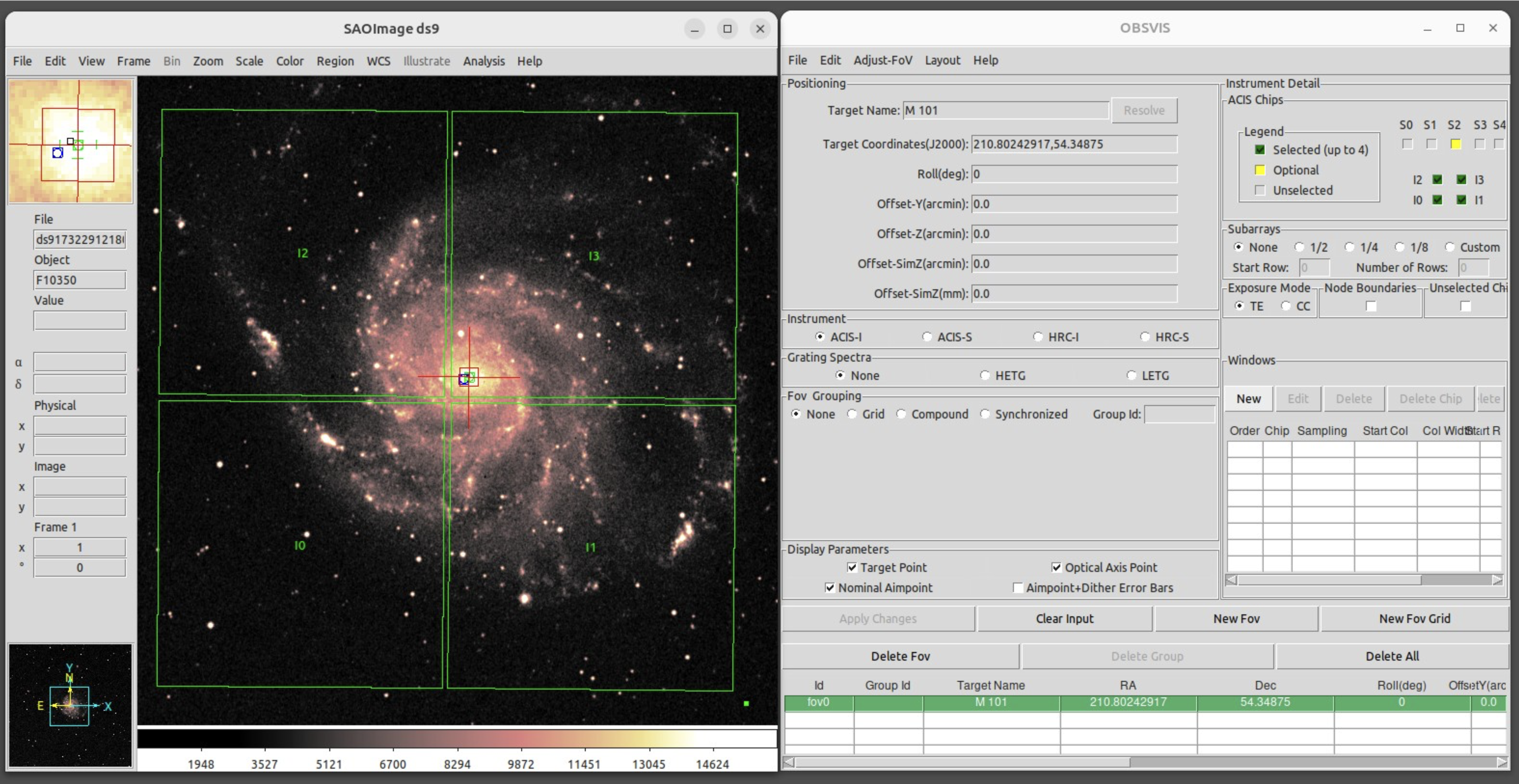}
\caption{Chandra Observation Visualizer (ObsVis) interface. The tool utilizes DS9 internal Tcl/Tk interpreter to overlay instrument fields of view and pointings on an optical image of M101, facilitating dynamic observation planning and instrument setup.}
\label{fig:obsvis}
\end{figure}

DS9 supports Chandra operations and proposal preparations through the Chandra Observation Visualizer (ObsVis). ObsVis enables interactive observation planning by allowing the overlay of instrument fields of view and pointings on celestial data (see Figure~\ref{fig:obsvis}). Within the X-ray community, DS9 is particularly valued for its ability to operate directly on event-level data, allowing filtering in energy, time, and spatial dimensions. 

Finally DS9 multiwavelength capabilities enable X-ray data to be interpreted alongside images and contours from other observatories, providing the framework necessary for modern multi-instrument analysis.

\subsection{Global Adoption}

DS9 is one of the most widely used general-purpose visualization environment in astronomy, and it is used across a broad range of wavelengths and scientific domains. In optical and near-infrared astronomy, it was employed for the visual classification of galaxies in the Hubble CANDELS survey \citep{kartaltepe2015}, and more recently, within the James Webb Space Telescope (JWST) community to generate three-color images during commissioning and to mitigate instrumental effects in early-release science programs \citep{rigby2023, rigby2025}. DS9 has also been used in planetary science, including analysis of NEAR-Shoemaker images of asteroid Eros and contributed to the development and validation of FITS standards for planetary data \citep{durda2012,marmo2018}. 

\begin{figure}[ht!]
\includegraphics[width=0.47\textwidth]{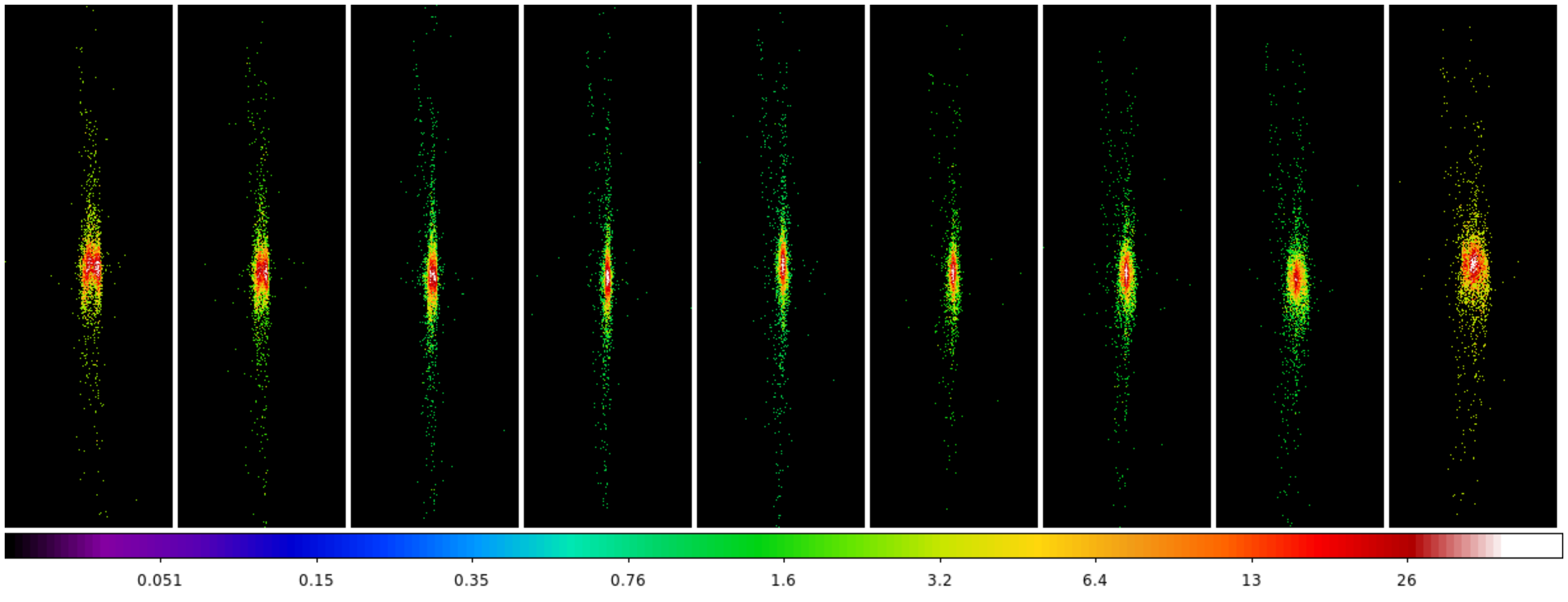}
\caption{DS9 multiframe-mode display of a focus search at the PANTER X-ray beamline. Each frame shows the measured X-ray point-spread function at a different detector position along the optical axis; comparing these images allows the best focus position to be determined. The data are sent to the DS9 display via XPA. Adapted from \citealt{burwitz2019}}
\label{fig:beamline}
\end{figure}

DS9 is widely utilized as a real-time display and quick-look analysis engine within observatory control systems and instrument pipelines. It serves as an operational standard for facilities including the W.M. Keck Observatory, the Dark Energy Survey (DES), and the Lick, Wendelstein, and Telescopio Nazionale Galileo observatories \citep{clarke2005, rossetti2006, kassis2016, Snigula2016, diehl2018}. It is used as the display system for European X-ray beamlines, including PANTER for the Advanced Telescope for High Energy Astrophysics (Athena) and Arcus optics characterization \citep{burwitz2019} (Figure~\ref{fig:beamline}), and XPBF and MINERVA for ongoing NewAthena calibrations \citep{heinis2024, krumrey2024}.

\begin{figure}[ht!]
\includegraphics[width=0.47\textwidth]{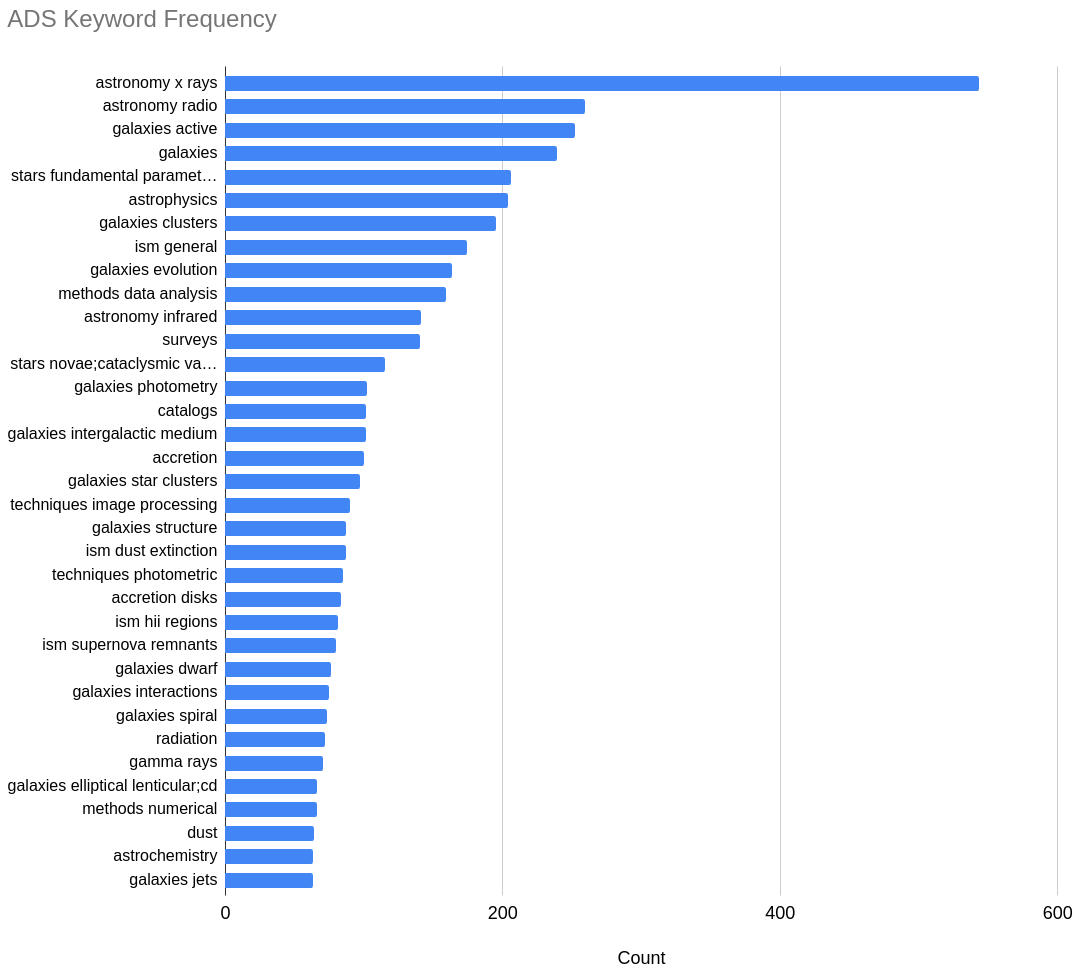}
\caption{Distribution of DS9 usage across research topics based on literature mentions. While X-ray astronomy accounts for the majority, the broad spread across radio, galaxies, and general astrophysics highlights DS9 role as a subject- and wavelength-independent tool.}
\label{fig:science}
\end{figure}

The distribution of DS9 usage across astrophysical domains (Fig.~\ref{fig:science}) reflects this breadth. While X-ray astronomy remains the largest category, studies of galaxies, radio astronomy, and active galactic nuclei also account for a large fraction of literature mentions.  Beyond research, DS9 is used in education and public engagement, serving as a core component of the NASA/IPAC NITARP teacher initiative, the "Analyzing the Universe" curriculum, and citizen science projects like the Hubble Asteroid Hunter \citep{matilsky2004, garvin2022}.

\begin{figure}[ht!]
\includegraphics[width=0.47\textwidth]{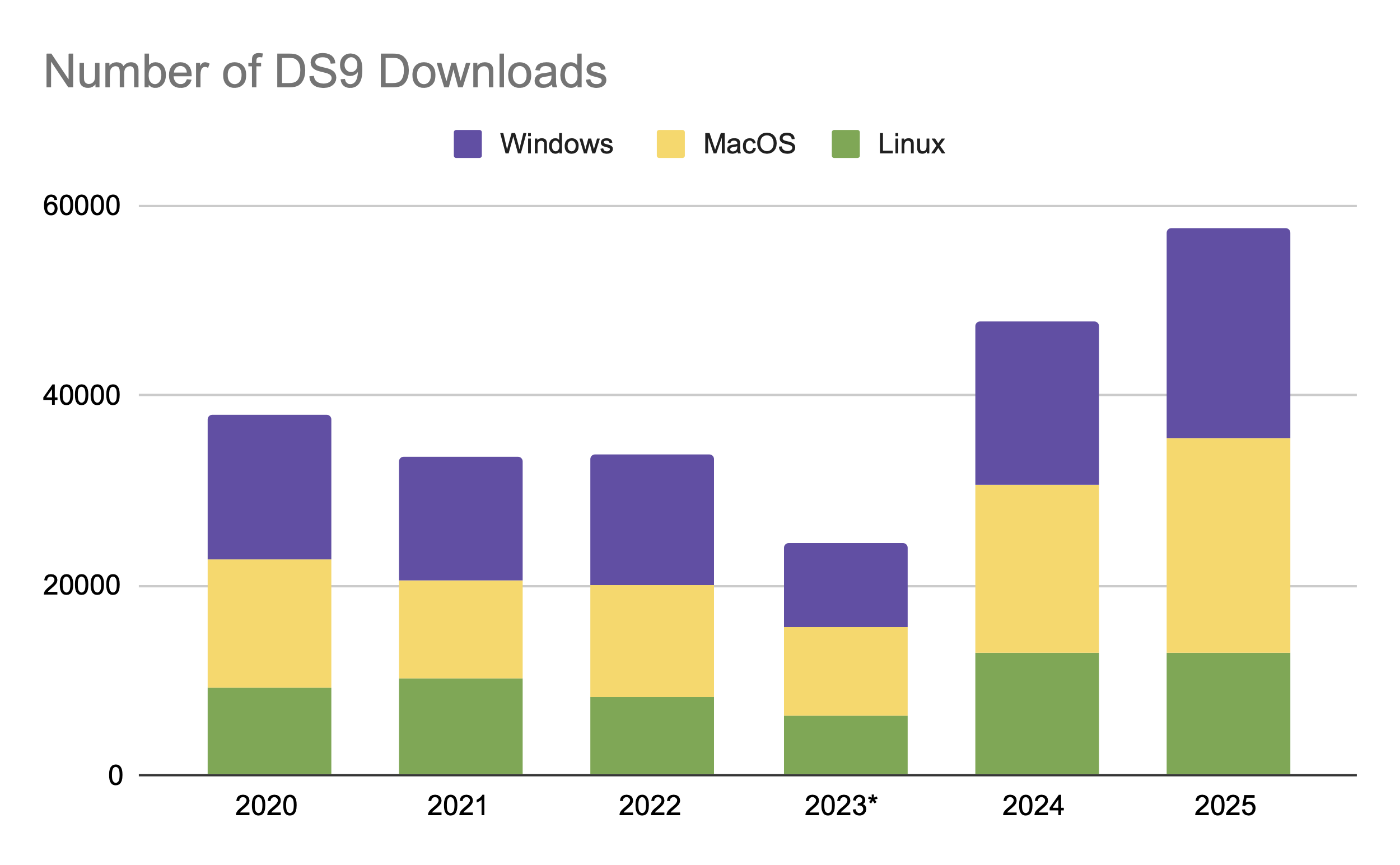}
\caption{
Number of DS9 downloads by operating system from 2020 to 2025. Downloads remained steady from 2020 to 2022 and increased substantially in 2024 and 2025. Windows accounts for the largest share in most years, followed by macOS and Linux.
Note: 2023 data are incomplete because of partial logging.
}
\label{fig:download}
\end{figure}

\begin{figure}[ht!]
\includegraphics[width=0.47\textwidth]{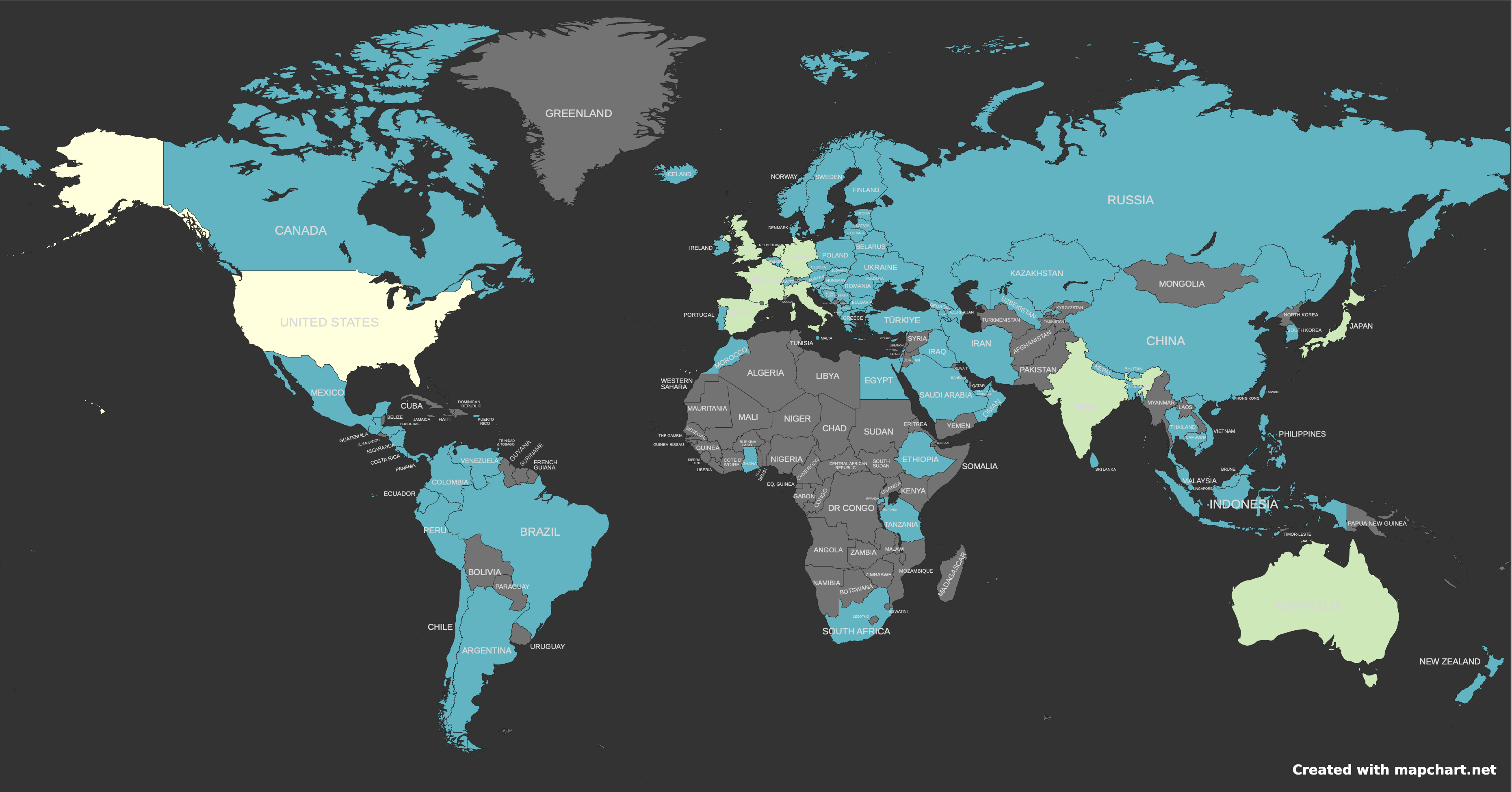}
\caption{Geographic distribution of downloads (Jan--Mar 2024). The United States (pale yellow) accounts for approximately 35\% of downloads, with the remaining $\sim$65\% distributed across Europe, Asia, Australia, and other regions. The next nine most active countries (pale green) each contribute 2.9--6\% of downloads, while additional downloads are distributed across many other countries (blue), demonstrating widespread international use. Gray indicates no recorded downloads.}
\label{fig:world}
\end{figure}

\begin{figure}[ht!]
\includegraphics[width=0.47\textwidth]{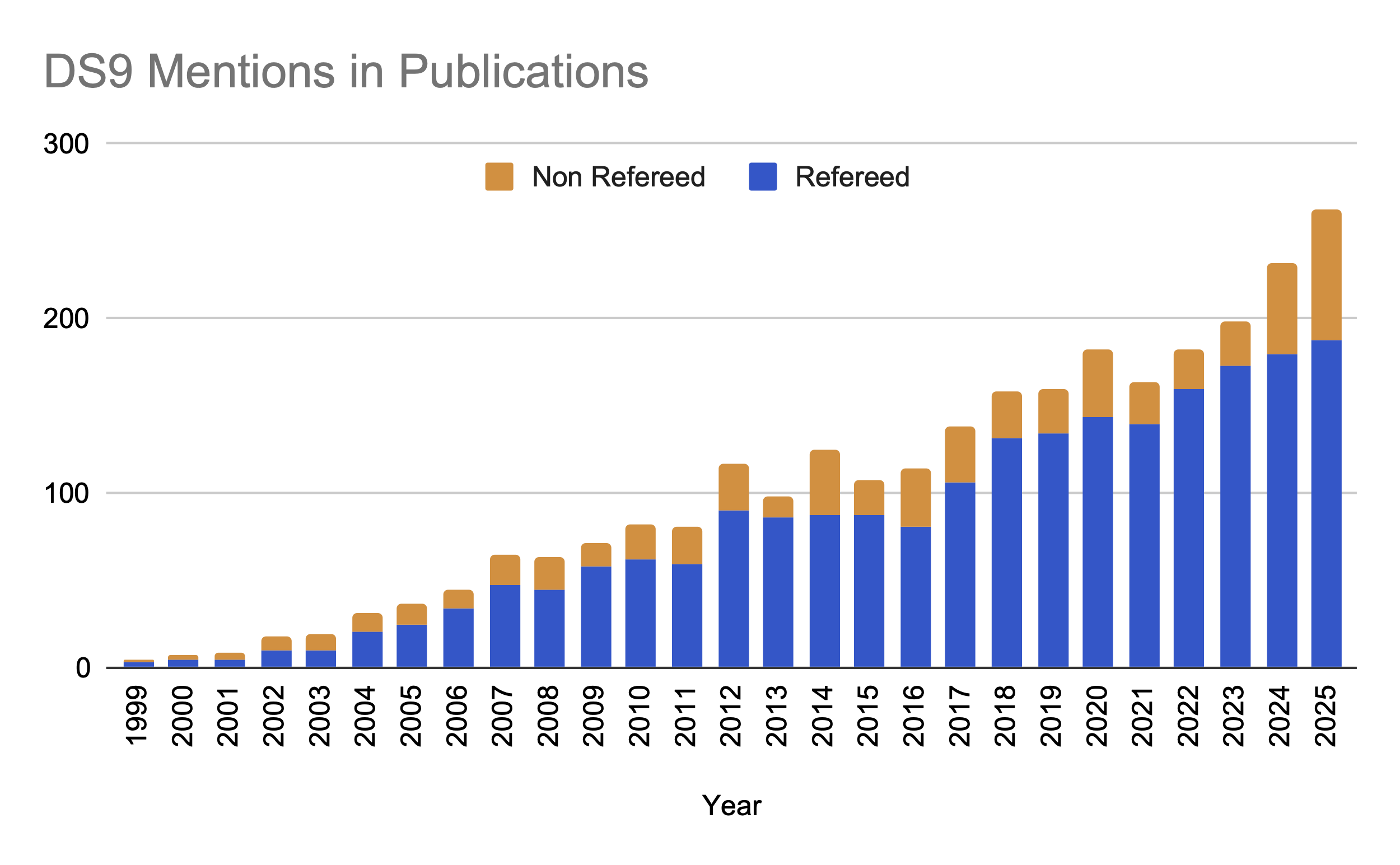}
\caption{Astrophysical Data System (ADS) search results for DS9 mentions in full-text publications (1999–2025). The steady growth in refereed publications provides a conservative lower limit on the software's impact, as software tools are frequently under-cited \citep{bouquin2020}.
}
\label{fig:publications}
\end{figure}

Download statistics show DS9’s broad use: from 2020 to 2025, DS9 was downloaded tens of thousands of times per year (Figure~\ref{fig:download}). The downloads span all major platforms, with Windows representing about 40--45\%, macOS about 35\%, and Linux about 25\%. Geographically, while the United States account for 35\% of users, the remaining 65\% are distributed worldwide (Figure~\ref{fig:world}). This broad adoption is also reflected in the steady growth of literature mentions, although part of the increase may reflect better awareness of the importance of citing software (Figure~\ref{fig:publications}). Because software remains historically under-cited in astronomical literature \citep{bouquin2020}, these metrics likely provide a lower limit on DS9's actual impact.

\section{Conclusion}

DS9 has become a fundamental tool in astronomy because it combines accurate astronomical coordinate handling, interactive performance, and cross-platform portability. Its longevity is a result of a modular architecture that has allowed it to adapt to evolving standards, larger and more complex datasets, and the emergence of the VO. DS9 is not simply a successful software package from the past, it is a mature and evolving environment that continues to support mission operations, professional research, and global education.

\section{Future Directions}
Despite its success, the long history of DS9 presents significant challenges for the maintenance of a mature software system.  The current architecture reflects decades of incremental updates and relies on legacy dependencies that are increasingly difficult to maintain within modern computing environments. While development is currently supported by the Chandra X-ray Center, the long-term outlook is linked to the life of the mission, which has now surpassed 25 years.
A DS9 modernization must preserve its core strengths of portability, rigorous coordinate handling, interoperability, and support for both interactive and scripted analysis, while transitioning to a more sustainable codebase.
Several broader computing trends also motivate modernization, for example the shift in display systems on Linux (from X11 to Wayland) and the growth of browser- and cloud-based analysis, where users expect visualization and analysis components to operate without a traditional local desktop application.

While no formal development has yet begun, concepts such as a future SAOImageDiscovery (a potential successor to DS9) represent a viable path forward for rethinking the platform. Securing the future of DS9 will require a sustained investment and commitment from the community to preserve the scientific reliability, transparency, and reproducibility that are essential for astronomical analysis. AI-assisted code review, automated refactoring, and dependency modernization may help determine whether DS9 can be updated incrementally or needs a more substantial redesign. Ultimately, the goal is to keep DS9 scientifically reliable and useful as astronomical datasets continue to grow in size and complexity.

\section{Acknowledging or Citing DS9}
\label{sec:cite}

If DS9 contributes to a scientific publication, we request that authors cite both the current  paper (A. Fruscione et al. 2026)---to acknowledge the software package and its ongoing development---and the original DS9 reference \citep{joye2003}---to acknowledge the introduction of DS9 to the astronomical community.

For AAS journal articles, DS9 should be included in the  \verb|\software{}| tag.

Authors may also include the following acknowledgment statement:
\begin{quote}
This research made use of SAOImageDS9, an astronomical imaging and data visualization application developed by the Smithsonian Astrophysical Observatory (A. Fruscione et. al 2026, \citealt{joye2003}).
\end{quote}


\begin{acknowledgments}
The authors gratefully acknowledge Michael Van Hilst and Eric Mandel, whose pioneering work on SAOImage, SAOtng, and the underlying software elements, forms the foundation on which DS9 was built. We also acknowledge the late Steve Murray for his leadership, vision, sustained support and funding of the SAOImage software tools that enabled their development. 

We thank the Astronomical Data Analysis Software and Systems (ADASS) community for awarding the 2024 Software Prize to William Joye and Eric Mandel ``in recognition of the outstanding contribution of SAOImageDS9 to the astronomical software community and the positive impact it has on many astronomy projects and scientists.''

Finally, we acknowledge the broader astronomical community, whose feedback, use, and contributions over many years have helped shape DS9 into the powerful and flexible tool it is today.

This work was supported by NASA contract NAS8-03060 to the CXC, which is operated by SAO for and on behalf of NASA. 

This research has made extensive use of the Astrophysics Data System, funded by NASA under Cooperative Agreement 80NSSC21M00561.

The authors thank the anonymous referee for carefully reading the manuscript and for providing insightful comments.

Generative AI tools were used during the preparation of this manuscript for editing, LaTeX formatting, and consistency checks. The authors maintain full responsibility for the scientific content and the accuracy of the final text.
\end{acknowledgments}

\begin{contribution}
A.F. was responsible for writing the full draft of the paper and submitting the manuscript, and contributed scientific input and feedback during the development of DS9. 

K.G. currently supports DS9 within the Chandra X-ray Center (CXC) and CIAO environments; he developed the Advanced View and DAX. All original figures in this paper were produced by him.  He is the author of a technical document summarized in Appendix A. 

W.J., now retired, initiated the development of DS9 in the late 1990s and was its primary developer throughout its history until his retirement in April 2026.

J.C.M., now retired, supervised the DS9 development over the past $\sim$10 years and authored the extensive \textit{``SAOImageDS9 Interface and Internal Design''} document, which forms the basis of Appendix B.

All authors provided technical feedback, critical revisions, and oversight for the paper.
\end{contribution}
%

\facilities{Chandra X-Ray Observatory, XMM-Newton, JWST, HST, Keck, Planck, IRSA, NEAR-Shoemaker, DART, Lick, Wendelstein Observatory, Telescopio Nazionale Galileo, Athena, Arcus, NewAthena, ALBA Synchrotron, PANTER, XPBF}


\software{SAOImageDS9, CIAO, Sherpa, IRAF, XPA, SAMP, WCS, HEALPix, Tcl/Tk, Python, AST, OpenSSL}



\appendix
\vspace{-1.5em}
\section{Code Architecture}
This appendix summarizes the technical design and implementation of the DS9 software system, including its architecture, codebase organization, and image rendering workflow.\\

\subsection{Architecture}
The code is implemented primarily in C/C++ for performance-critical components and in Tcl/Tk for the graphical user interface and scripting layer.  The Tk event loop drives the application, managing both rendering and user interaction. 

\begin{figure}[h!]
\centering
\includegraphics[width=0.8\textwidth]{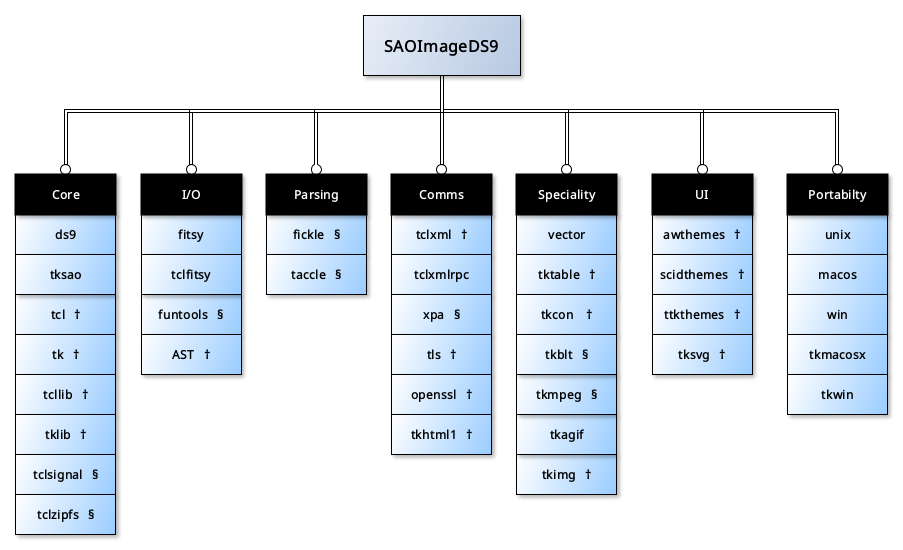}
\caption{Functional grouping of DS9 components. Packages without annotation are original code developed specifically to support DS9. Packages listed with $\dagger$ are developed and maintained by third parties. Packages listed with \S \, were originally developed by third parties, but have since been orphaned; forks for these packages are now maintained to support DS9. }
\label{fig:packages}
\end{figure}

DS9 is a modular system composed of multiple packages (see Fig.~\ref{fig:packages}). The DS9 project develops and maintains 20 packages and uses an additional 16 third-party packages. Over time, several critical components, such as \texttt{funtools}, \texttt{xpa}, and \texttt{tkblt}, have been forked into the DS9 project after being abandoned upstream, as they remain essential for core functionality with no viable replacements.

The DS9 code is organized into packages that handle distinct functional domains. The graphical interface is implemented through \texttt{ds9} and \texttt{tksao}, while data handling relies on \texttt{fitsy} and \texttt{funtools}. Plotting capabilities are provided by \texttt{tkblt}, communication is handled by \texttt{xpa}, and supporting utilities include libraries such as \texttt{vector} and \texttt{tclsignal}. Structured input parsing is handled via flex\footnote{\url{https://en.wikipedia.org/wiki/Flex_(lexical_analyzer_generator)}} and bison\footnote{\url{https://en.wikipedia.org/wiki/GNU_Bison}} equivalent parsers generated with \texttt{fickle} and \texttt{taccle}.

Additional Tcl/Tk extensions provide support for XML-RPC (Remote Procedure Call) communication (\texttt{tclxmlrpc}) communication used for SAMP messages, in-memory file systems (ZIP-based, \texttt{tclzipfs}), animation and MPEG generation (\texttt{tkagif} and \texttt{tkmpeg}), as well as platform-specific utilities. The system also depends on several major third-party libraries, including the AST library for WCS and FITS coordinate handling, OpenSSL for secure communication, and the Tcl/Tk core libraries along with their associated extensions.

\subsection{Graphical User Interface}

\begin{figure}[h]
\centering
\includegraphics[width=0.8\textwidth]{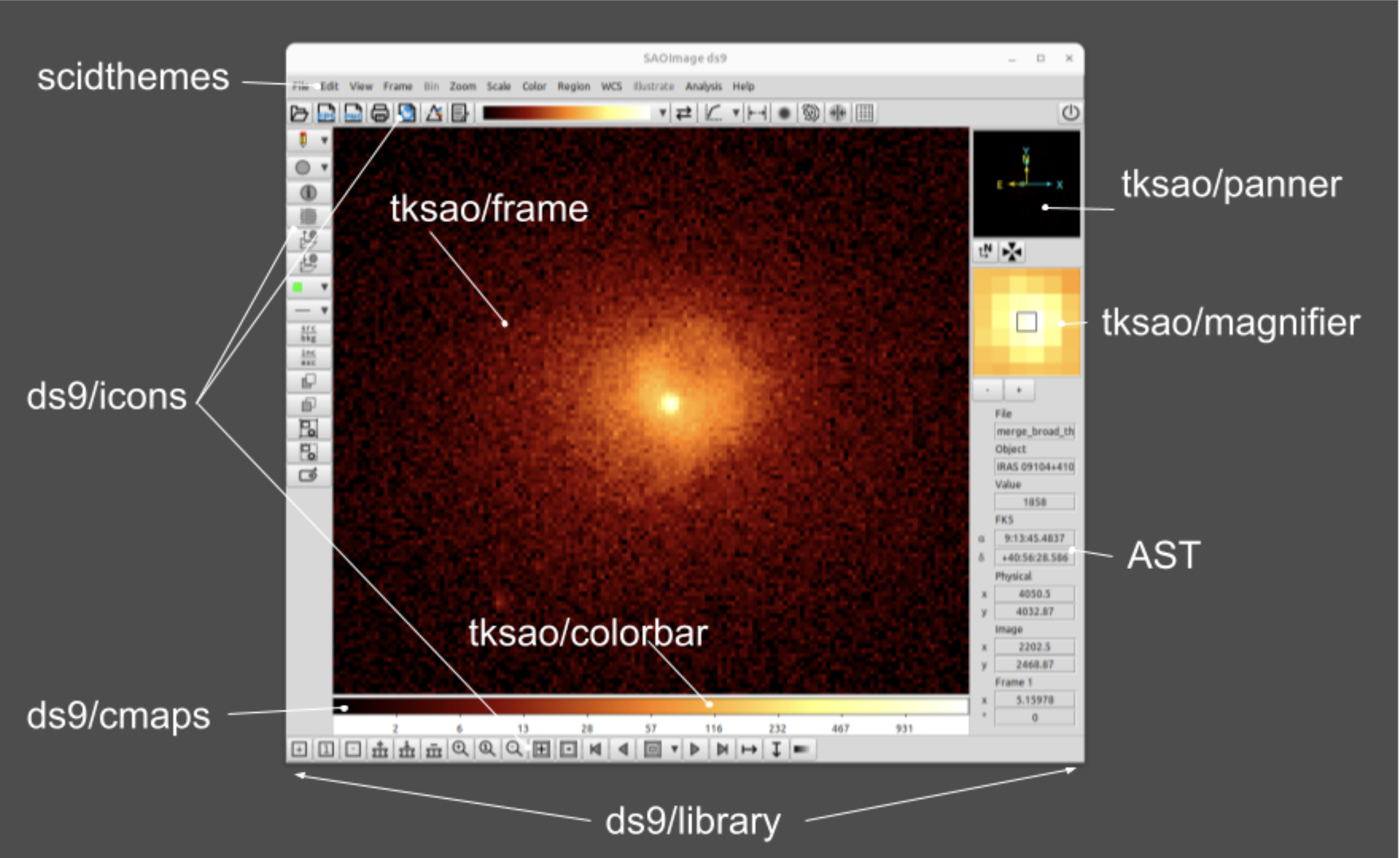}
\caption{The DS9 user interface with certain components labeled with their code source; not all components are labeled. The Tcl/Tk ds9/library module provides the top level window in which all the other elements are displayed. In this example the "Advanced" view is being used. This module also handles all the events such as clicking to create new regions or panning the frame. The image itself is a custom frame widget written in C++. Similarly the magnifier, panner, and colorbar are all custom widgets written in C++.  The AST library is used to compute the celestial coordinates used in the information panel. The overall look and feel of the user interface (font color, size, style, spacing, etc) is provided by the base "ttk" widget classes; the "scidthemes" provides a collection of styles.
}
\label{fig:gui}
\end{figure}

The structure of the interface is illustrated in Fig.~\ref{fig:gui}.
The GUI consists of a top-level window managed by the \texttt{ds9/library} module, with custom C++ widgets used for image display, magnification, panning, and colorbar visualization. Celestial coordinate computations are performed using the AST library.

\subsection{Repository}

DS9 is distributed as open-source software and hosted on GitHub \footnote{\url{https://github.com/SAOImageDS9/SAOImageDS9}}. The repository uses submodules to track specific versions of third-party dependencies, ensuring reproducibility across builds. Although the project is open to external contributions, community input has historically been limited. Several third party distributions also exist in certain Linux repositories (e.g. Debian and Arch), as well as MacPorts for macOS. 

\subsection{Packaging and Distribution}

A single codebase is used to support multiple windowing systems, including X11 on Linux, macOS (both X11 and Aqua), and Windows via Cygwin. The build system is based on \texttt{autoconf} and \texttt{automake}, producing self-contained binaries that include all required dependencies.

Platform-specific packaging strategies vary: Linux distributions are provided as self-decompressing binaries with embedded resources; macOS builds are distributed either as DMG installers or as binaries with separate resource archives; and Windows versions are packaged as self-extracting executables. DS9 as of version 8.7 provides binaries for approximately 20 platforms.

\subsection{Implementation Languages}

The DS9 repository includes a mix of programming and markup languages, dominated by C/C++ (50.4\%) and Tcl/Tk (19.9\%) for the GUI and scripting. Additional languages include HTML and TeX for documentation, Roff for manual pages, and M4 for build configuration which make up the remaining 29.7\%. 


\section{Interface and Internal Design Overview}

This appendix provides a technical summary of the SAOImageDS9 software, based on the comprehensive documentation \textit{``SAOImageDS9 Interface and Internal Design''}\footnote{\url{https://ds9.si.edu/reference/ds9.pdf}}. The goal is to highlight the key architectural components, data handling mechanisms, and visualization pipeline relevant for scientific analysis, while preserving the essential technical details of the original document.

\subsection{Control Architecture}

DS9 can be controlled in three ways. Normally, the application is controlled by the GUI menus. DS9 also has a command language that can be used from the Unix command line with XPA commands, setting DS9 parameters and returning them:
\begin{verbatim}
xpaset -p ds9 [command] [value]
xpaget ds9 [command]
\end{verbatim}
The same commands can also be used on the initial DS9 startup command line, where they are executed in sequence from left to right. XPA commands may be issued one at a time from the terminal or collected in a script to run in batch mode. Commands may be sent to DS9 as long as the GUI is open. DS9 can also be commanded on one machine using XPA commands sent from another machine.

\subsection{FITS Handling}

When a FITS file is opened successfully, DS9 displays a subset of its main image, specifically the first FITS HDU containing a non-null image. FITS files can contain multiple segments, or HDUs, and DS9 examines the HDUs in turn to identify valid image data. A primary HDU is selected if it contains a non-null image; other HDUs may also be selected if they contain a non-null image, a compressed image, an event file, or a HEALPIX image.

The FITS \texttt{OBJECT} and \texttt{UNITS} keywords are used to fill the corresponding entries in the information panel. The \texttt{BSCALE} and \texttt{BZERO} keywords are used to scale the data as it is read in. The \texttt{BLANK} keyword defines a null value, and the \texttt{DATASEC} keyword defines a file subsection. By default, if the \texttt{DATASEC} keyword is present, DS9 uses it to determine what subset of the data is considered valid for min/max calculations and display.

An X-ray event file is stored as a FITS table with columns usually including \texttt{X} and \texttt{Y}. DS9 recognizes an event file when it finds a binary table with \texttt{EXTNAME} equal to \texttt{EVENTS}, \texttt{STDEVENT}, or \texttt{RAYEVENT}, and with column names \texttt{X} and \texttt{Y} present. When the file is loaded, DS9 makes an image on the fly by binning the data on the \texttt{X} and \texttt{Y} columns. Columns with other names can also be used as the columns to bin on.

\subsection{Rendering Pipeline}

DS9 supports a number of operations on the displayed image, including binning or blocking, cropping, smoothing, scaling, color mapping, zooming, and panning. For table data, DS9 first determines whether the table is an event file. Event data are binned, while other table data are decompressed. Image data proceed directly to blocking. The subsequent operations are then applied in the following order:
\begin{center}
\textit{blocking} $\rightarrow$ \textit{cropping} $\rightarrow$ \textit{smoothing} $\rightarrow$ \textit{scaling} $\rightarrow$ \textit{color mapping} $\rightarrow$ \textit{zooming}
\end{center}
For event files, the binning step precedes blocking:
\begin{center}
\textit{binning} $\rightarrow$ \textit{blocking} $\rightarrow$ \textit{cropping} $\rightarrow$ \textit{smoothing} $\rightarrow$ \textit{scaling} $\rightarrow$ \textit{color mapping} $\rightarrow$ \textit{zooming}
\end{center}
DS9 also distinguishes between the full image, the part of the image currently in memory, the part displayed in the selected display frame, and the part displayed in the magnifier.




\subsection{Visualization and Scaling}

DS9 maps image pixel values to displayed RGB values using scale limits, a scaling function, color stretch, bias, and a color table. The scale limits $x_1$ and $x_2$ define the range of pixel values to be displayed. Pixel values are first converted to a normalized value used by the selected scaling function.

DS9 provides linear, logarithmic, power, square root, squared, arc-hyperbolic sine, hyperbolic sine, and histogram equalization functions. For example, the logarithmic and arc-hyperbolic sine scaling functions are
\begin{align}
s(r) &= \frac{\log(pr+1)}{\log(p)}, \\
s(r) &= \frac{\sinh^{-1}(pr)}{3}.
\end{align}

After scaling, DS9 applies the color stretch and bias before using the color table lookup to determine the displayed RGB value. This mapping is represented by
\begin{equation}
F(x)=c(y),
\end{equation}
with
\begin{equation}
y=N_1\left(a s(r)+t\right)
\end{equation}
and
\begin{equation}
t=\frac{1}{2}\left(1-a(b+0.5)\right),
\end{equation}
where $a$ is the color stretch, $b$ is the bias, and $c(y)$ is the color table lookup.

DS9 allows the user to specify the scale limits directly or to compute them using algorithms in the Scale menu. The ZScale algorithm, inherited from IRAF, samples the pixel values, sorts them into a monotonic array, determines the median value, fits a line to the sorted values, and uses the fitted slope together with the contrast parameter to estimate the lower and upper scale limits.

\subsection{Coordinate Systems}

DS9 supports coordinate conversion among image, physical, and WCS coordinate systems. WCS coordinates are defined in the FITS header as a parameterized mapping from image physical coordinates. DS9 uses the AST library to compute celestial coordinates and supports transformations among standard astronomical coordinate systems, including FK4, FK5, ICRS, Galactic, and Ecliptic coordinates.

\subsection{Regions and Spatial Analysis}
DS9 regions are geometric shapes used to mark or select areas in an image. Supported region shapes include circles, annuli, ellipses, boxes, polygons, lines, points, and text. Regions may be defined in image, physical, or WCS coordinates, and can be edited interactively or loaded from region files.
Regions can be used for measurements and analysis within selected areas, including counts and statistics. DS9 also supports Boolean operations on regions, allowing regions to be included, excluded, or combined to form more complex selections.

\subsection{Advanced Data Handling}

DS9 supports data cubes and event files. Event files are stored as FITS tables and can be displayed as images by binning selected columns, such as the \texttt{X} and \texttt{Y} columns. DS9 also supports multiple frames and RGB images for displaying more than one image or image component. Additional operations, including contour generation and smoothing, are supported.

\subsection{Performance}

DS9 uses memory mapping to support large image files, allowing image data to be accessed without reading the entire file into memory at once. This permits DS9 to display and process images that are larger than the available physical memory.

Certain operations in DS9 are multi-threaded, including 3D rendering, smoothing, and contour generation. These operations can make use of multiple processors and may reduce processing time depending on the number of processors available.

\subsection{Interoperability}

DS9 interfaces with external tools through XPA, SAMP, and Tcl scripting. XPA provides interprocess communication through commands such as \texttt{xpaset} and \texttt{xpaget}. SAMP support follows the VO SAMP standard and allows DS9 to communicate with other applications. DS9 can also be controlled from Astropy through SAMP, and it provides interfaces used with software such as CIAO and IRAF.

\subsection{Customization}
DS9 is implemented using Tcl/Tk. Visualization relies on color table lookup arrays for mapping scaled pixel values to RGB values. User customization is supported through startup and preference files that allow persistent configuration of DS9 behavior.
\bibliography{ds9bibliography}{}
\bibliographystyle{aasjournalv7}



\end{document}